%% file: main.tex
\documentclass[aps,prl,reprint,superscriptaddress]{revtex4-2}

\pdfoutput=1 %% Uncomment for arxiv Submission

\usepackage{amsmath}
\usepackage{amssymb}
\usepackage{bm}
\usepackage{physics}

\usepackage{hyperref}
\usepackage[capitalize]{cleveref}

\usepackage[usenames]{xcolor}
\usepackage{graphicx}

\DeclareMathOperator{\arsinh}{arsinh}
\newcommand{\ii}{\mathrm{i}}

\begin{document}

\title{Fluctuation-dominated quantum oscillations in excitonic insulators}

\author{Andrew A. Allocca} \email{aallocca@lsu.edu}
\affiliation{Department of Physics and Astronomy and Center for Computation and Technology, Louisiana State University, Baton Rouge, LA 70803, USA}
\affiliation{T.C.M. Group, Cavendish Laboratory, University of Cambridge, JJ Thomson Avenue, Cambridge, CB3 0HE, U.K.\looseness=-1}
\author{Nigel R. Cooper}
\affiliation{T.C.M. Group, Cavendish Laboratory, University of Cambridge, JJ Thomson Avenue, Cambridge, CB3 0HE, U.K.\looseness=-1}
\affiliation{Department of Physics and Astronomy, University of Florence, Via G. Sansone 1, 50019 Sesto Fiorentino, Italy\looseness=-1}

\date{\today}

\begin{abstract}
The realization of excitonic insulators in transition metal dichalcogenide systems has opened the door to explorations of the exotic properties such a state exhibits. 
We study theoretically the potential for excitonic insulators to show an anomalous form of quantum oscillations: the de Haas-van Alphen effect in an insulating system. 
We focus on the role of the interactions that generate the energy gap and show that it is crucial to consider quantum fluctuations that go beyond the mean field treatment. 
Remarkably, quantum fluctuations can be dominant, and lead to quantum oscillations than are significantly larger than those predicted using mean field theory. 
Indeed, in experimentally accessible parameter regimes these fluctuation-generated quantum oscillations can even be larger than what would be found for the corresponding gapless system.
\end{abstract}

\maketitle

Materials that become insulating as the result of interactions have attracted significant interest in recent years.
Excitonic insulators, formed by the spontaneous condensation of electron-hole bound states, were theoretically proposed more than 50 years ago~\cite{Cloizeaux1965,Jerome1967,Keldysh1968,Halperin1968}. 
They have now been realized in single~\cite{Jia2022} double layer~\cite{Wang2019,Ma2021} transition metal dichalcogenide (TMD) systems, a class of materials which itself has garnered wide and significant attention~\cite{Manzeli2017}. 
Kondo insulators~\cite{Menth1969,Hewson1993}, in particular topological Kondo insulators~\cite{Dzero2010,Dzero2016}, have also been intensely studied because of their nontrivial topological properties and significant bulk gaps at low temperature.

The measurement of oscillations of the  magnetization with magnetic field -- {\it i.e.} the de Haas-van Alphen effect~\cite{deHaas1930} -- in the Kondo insulators SmB${}_6$~\cite{Li2014,Tan2015,Hartstein2018,Hartstein2020} and YbB${}_{12}$~\cite{Liu2018,Xiang2018,Liu2022} was particularly unexpected, since the presence of a Fermi surface was long believed to be a necessary condition to realize this effect~\cite{Lifshitz1956,Shoenberg1984}.
Consequently, a large amount of theoretical work has gone into understanding how quantum oscillations (QOs) can manifest in insulators, some focused specifically on these Kondo systems~\cite{Baskaran2015,Erten2016,Knolle2017a,Erten2017,Riseborough2017,Sodemann2018,Chowdhury2018,Peters2019,Lu2020,Varma2020}, and others considering more general insulating systems~\cite{Knolle2015,Zhang2016,Pal2016,Pal2017a,Pal2017b,Knolle2017b,Shen2018,Skinner2019,Lee2021,He2021,Allocca2022,Randeria2023,Julian2023}, including excitonic insulators which will be our focus here.

Direct calculations show that even simple models of non-interacting band insulators can exhibit QOs~\cite{Knolle2015,Zhang2016,Pal2016,Pal2017a,Pal2017b,Knolle2017b,Shen2018,He2021,Allocca2022,Randeria2023,Julian2023}: provided the minimum band-gap traces out some closed area in reciprocal space the free energy contains an oscillatory component and therefore so do thermodynamic quantities like the magnetization. 
The resulting QOs have several properties that do not depend on specific details of the model.  
First, the oscillation frequency is determined by the area noted above, just as the area of the Fermi surface determines this frequency in metals. 
Second, at small magnetic field $B$ these oscillations are suppressed by a factor of the form $\exp(-B_0/B)$, where $B_0$ is proportional to the size of the band gap.   
While this has the same form as the Dingle suppression of QOs in metals due to impurity scattering~\cite{Shoenberg1984}, we emphasize that here $B_0$ is an intrinsic property of the system. 
We have previously shown that -- at least for the lowest frequency oscillatory response -- these results extend beyond the case of non-interacting particles, and arise also for interaction-driven excitonic and Kondo insulators when the interactions are treated within mean-field theory~\cite{Allocca2022}. 

\begin{figure}
    \centering
    \includegraphics[width=\columnwidth]{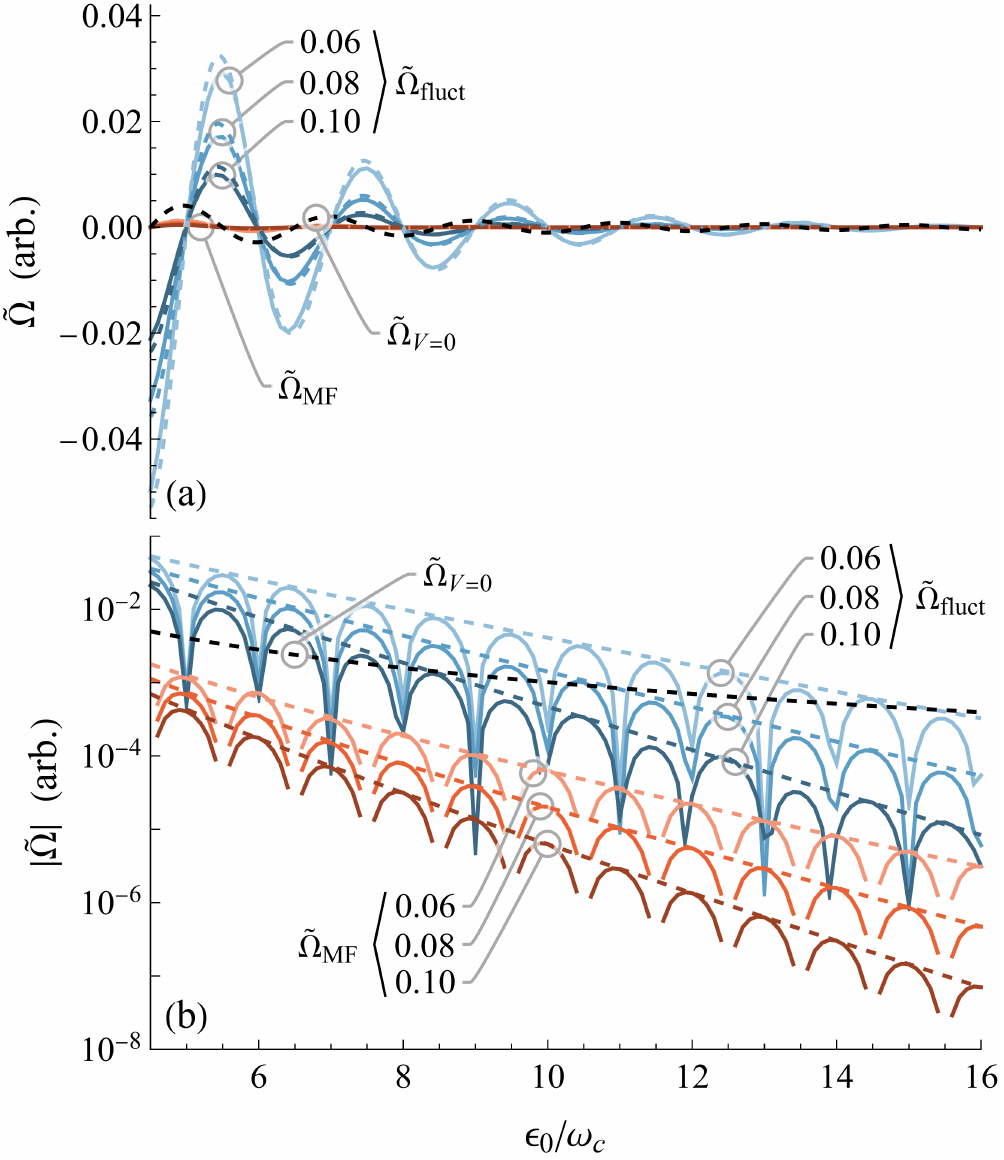}
    \caption{(Color Online) The oscillatory free energies $\tilde\Omega_\mathrm{fluct}$ (blue), $\tilde\Omega_\mathrm{MF}$ (red), and $\tilde\Omega_{V=0}$ (black) are plotted on (a) linear and (b) logarithmic scales as functions of inverse cyclotron frequency $(\propto 1/B)$.
    The numbers labeling different curves correspond to three choices of the size of the gap $\Delta_0$, given in units of $\epsilon_0$.
    Solid lines are the results of numerical evaluation and dashed lines of the same color are corresponding analytic results, \cref{eq:OmegaMFtilde,eq:Omegaflucttilde,eq:OmegaV=0}. 
    In (a) the three mean field results are too small to distinguish.
    In (b) only the amplitude of analytic results are plotted.}
    \label{fig:oscillations}
\end{figure}

In this paper we go beyond the mean field approximation for a model of an excitonic insulator, and calculate the contributions to QOs arising from the quantum fluctuations of the gap. 
We find the very surprising result that these quantum fluctuations give a contribution that dominates the QOs. 
Indeed, as shown in \cref{fig:oscillations}, we find that in experimentally accessible low-electron-density parameter regimes, for instance in TMD double layer systems~\cite{Wang2019,Ma2021}, the oscillations from fluctuations are significantly larger than those obtained from just a mean-field treatment of the interactions.
Even more strikingly, these oscillations can be of the same size or larger than those for the corresponding gapless system obtained by ``turning off'' the interaction. 
Counterintuitively, for low-density systems QOs can be amplified by introducing interactions that destroy the Fermi surface. 
We study the system in a regime where quantum fluctuations give a correction to the free energy $\Omega_\mathrm{fluct}$ that is much smaller than the total free energy $\Omega$, as is necessary for a stable mean field theory. 
Since the part of the free energy that oscillates with magnetic field, $\tilde\Omega$, is a very small fraction of the total free energy $\Omega$, there is no contradiction that the quantum fluctuations can be the dominant source of $\tilde\Omega$ while still being small compared to $\Omega$.

As we will explain below, we expect that this effect is very generic. 
It can be traced back to the contributions of electron-electron interactions to the energy offsets of the relevant bands.
However, in order to establish the importance of the effect most clearly, we study a concrete model for which we are able to provide an exact calculation of the leading order fluctuation effects.

We consider a two-dimensional, two-band system of spinless electrons with an interband interaction. 
For vanishing magnetic field it is described by the Hamiltonian%action
\begin{equation}
\hat{H} = \sum_\mathbf{k}\sum_{i=c,v} \xi_{i, \mathbf{k}} \hat{\psi}^\dagger _{i,\mathbf{k}}\hat{\psi}_{i,\mathbf{k}} + \frac{V}{A} \!\!\sum_{\mathbf{q},\mathbf{k},\mathbf{k}'}\!\! \hat{\psi}^\dagger_{c,\mathbf{k}+\mathbf{q}}\hat{\psi}^\dagger_{v,\mathbf{k}'-\mathbf{q}} \hat{\psi}_{v,\mathbf{k}'} \hat{\psi}_{c,\mathbf{k}}
\label{eq:ham}
\end{equation}
where $c,v$ label the conduction and valence bands and $A$ is the area of the system. 
The constant $V>0$ parameterizes the strength of the attractive Coulomb interaction between particles and holes, here approximated as a contact interaction. 
We consider a particle-hole symmetric model, with chemical potential $\mu=0$ and with dispersions $\xi_{c,\mathbf{k}} =  - \xi_{v,\mathbf{k}}  = \abs{\mathbf{k}}^2/2m  - \epsilon_0/2  \equiv\xi_k $.
This assumption simplifies our analysis but is not essential for arriving at our main result --- broken particle-hole symmetry should not have a significant qualitative effect. 
(We note, however, that TMD double layer systems that realize the sort of system we are interested in are well approximated as particle-hole symmetric.)
The energy $\epsilon_0>0$ is the offset of the two band edges. 
We also need to define a UV cutoff for $\xi_\mathbf{k}$, which we set at an energy $\Lambda$ from the band minimum so that $\mathrm{max}(\xi_k) = \Lambda - \epsilon_0/2$.
(In terms of real material parameters, this $\Lambda$ is related to the bandwidth.) 
Thus the total electron density in the system is $n_e = \rho_F \Lambda$, with densities $n_c = \rho_F \epsilon_0/2$ and $n_v = \rho_F (\Lambda - \epsilon_0/2)$ in the conduction and valence bands respectively, where $\rho_F = m/2\pi$ is the density of states for a spinless 2d electron gas.
Here and throughout we set $\hbar=c=1$.

\begin{figure}
    \centering
    \includegraphics[width=0.95\columnwidth]{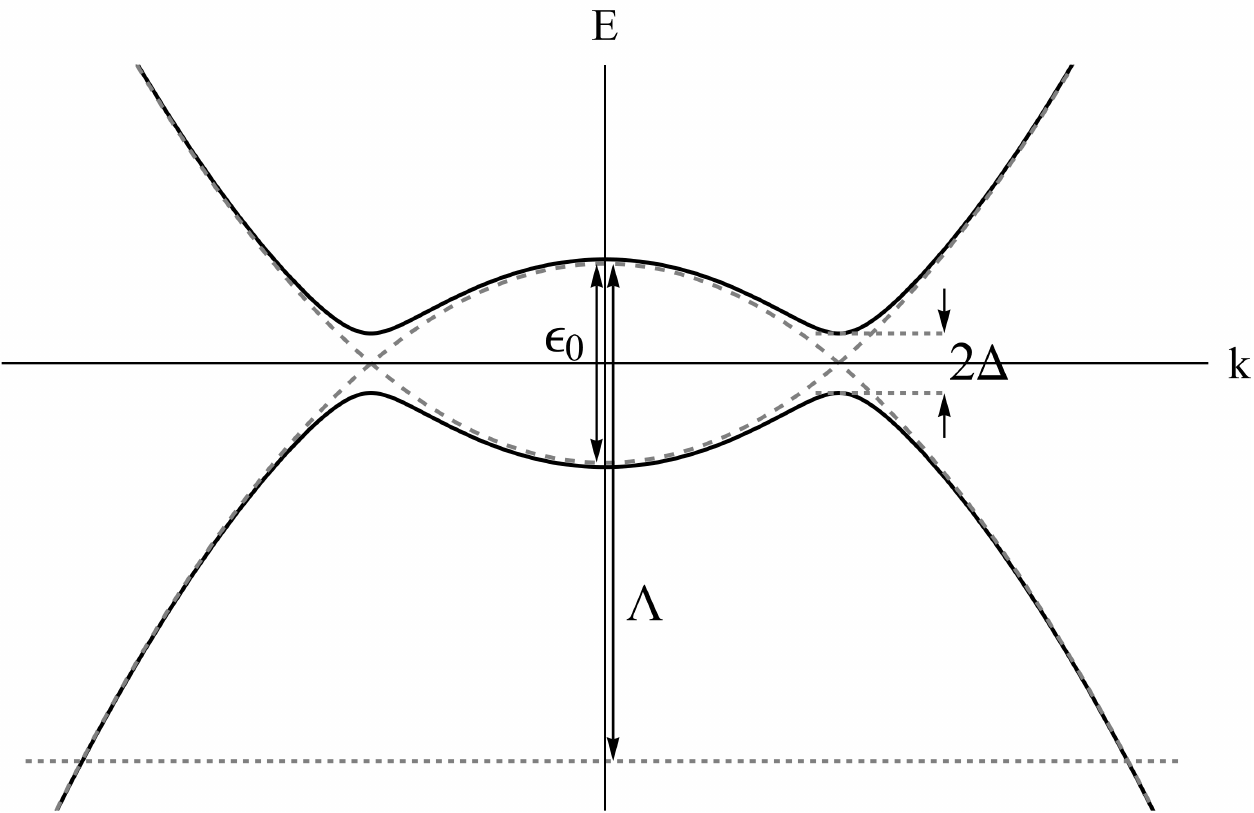}
    \caption{The particle-hole symmetric band structure we consider.
    The solid lines are $\pm E_k$ and the dashed lines are $\pm\xi_k$.
    The band gap $2\Delta$, band edge offset $\epsilon_0$, and UV cutoff $\Lambda$ in the valence band are indicated.}
    \label{fig:bands}
\end{figure}

A systematic analysis of the thermodynamics of this model is conveniently performed using standard finite-temperature field theoretical methods. 
We thus introduce Grassmann fields for the conduction and valence electrons,
$\psi_{c,k}$ and $\psi_{v,k}$, with the subscript $k$ representing both momentum $\mathbf{k}$ and Matsubara frequency $\epsilon_n = (2n+1)\pi/\beta$ at inverse temperature $\beta$. 
The interaction decouples with a Hubbard-Stratonovich transformation in terms of a bosonic field $\Delta_q$ related to the pairing of electrons and holes between the two bands. 
We separate $\Delta_q = \delta_{q,0}\Delta + \eta_q$ into a static, spatially uniform mean field $\Delta$ and a dynamic, spatially-nonuniform fluctuation field $\eta_q$.
Choosing $\Delta$ to be real, we identify the real and imaginary parts of $\eta_q$ as the Higgs (or amplitude) mode and phase mode respectively. 
The resulting action is $S = S_\mathrm{MF} + S_\mathrm{fluct}$ with
\begin{gather} 
    S_\mathrm{MF} = \frac{\beta A}{V}\Delta^2 + \sum_k \bar{\Psi}_k \mqty(-\ii\epsilon_n + \xi_\mathbf{k} & -\Delta \\ -\Delta & -\ii\epsilon_n - \xi_\mathbf{k})\Psi_k \label{eq:SMF0}\\
    S_\mathrm{fluct} = \frac{\beta A}{ V}\sum_q \bar{\eta}_q \eta_q - \sum_{k,q} \bar{\Psi}_{k+\tfrac{q}{2}}\, \mqty(0 & \eta_q \\ \bar{\eta}_{-q} & 0)\Psi_{k-\tfrac{q}{2}}, \label{eq:Seta0}
\end{gather}
where $\Psi_k = (\psi_{c,k} \quad \psi_{v,k})^T$. 
$S_\mathrm{MF}$ is the mean field action for the electrons, in which $\Delta$ is determined self-consistently to minimize the free energy. 
Diagonalizing the mean-field single-particle Hamiltonian yields the new set of gapped bands, $\pm E_k = \pm\sqrt{\xi_k^2+\Delta^2}$.
These and the relevant energies of the system are shown in \cref{fig:bands}.

The action $S_\mathrm{fluct}$ describes the effects of quantum fluctuations beyond mean field theory. 
The consequences of these fluctuations for the thermodynamics of related BCS superconductors have been worked out in detail previously~\cite{Vaks1962,Kos2004,Hoyer2018}. 
Much can be gleaned from these studies for our $B=0$ model of the excitonic insulator \cref{eq:ham}, which is equivalent to a superconductor under a particle-hole transformation.

Here we apply this same approach to the excitonic insulator subjected to an external magnetic field $B$, as obtained by minimal coupling to a vector potential. 
The modified kinetic energy term in \cref{eq:ham} is then diagonal in the basis of Landau level states, with energies $\xi_l = \omega_c(l+1/2) -\epsilon_0/2$ that are evenly spaced by the cyclotron energy $\omega_c = eB/m$.  
Expanding the contact interaction \cref{eq:ham} in the Landau level basis shows that it mixes states in different Landau levels. 
We are still able to perform all equivalent steps of the analysis that led to \cref{eq:SMF0,eq:Seta0}, but we now arrive at new field-dependent action given in the Supplement~\cite{Supplement}. 
As well as changing the electronic states into highly degenerate Landau levels, the nonzero magnetic field also generates an additional $B$-dependent factor in the second term of \cref{eq:Seta0} describing the coupling between electrons and the fluctuations. 
Since the gap must be determined self-consistently it also acquires a field dependence, $\Delta\to\Delta(B)$.

The free energy is obtained by integrating out all fields, both electrons and fluctuations. 
This leads to a total free energy $\Omega(B)$ that is the sum of mean-field $\Omega_\mathrm{MF}(B)$ and fluctuation $\Omega_\mathrm{fluct}(B)$ contributions. 
Despite the nonzero magnetic field, the fluctuation free energy can be written in the same form as for $B=0$~\cite{Vaks1962,Kos2004,Hoyer2018},
\begin{equation}\label{eq:Omegafluct}
    \Omega_\mathrm{fluct}(B) = \frac{1}{2\beta}\sum_q \tr\ln\left(\hat{\mathbf{1}}+V\hat{\Pi}_q\right).
\end{equation} 
The polarization $\hat\Pi_q$ (provided in the Supplement~\cite{Supplement}) contains all information about Landau quantization from the external field. 
It has features not seen for $B=0$, such as a coupling of the Higgs and phase modes. 

We are interested in the oscillatory components of these thermodynamics quantities as a function of the (inverse) magnetic field.
We shall only consider regimes of weak magnetic fields where all of $\Delta(B)$,  $\Omega_\mathrm{MF}(B)$, and $\Omega_\mathrm{fluct}(B)$ have oscillation amplitudes that are  small compared to their zero-field values. 
We thus separate these into their non-oscillatory values taken for $B=0$ -- denoted $\Delta_0$, $\Omega_{\mathrm{MF},0}$ and $\Omega_{\mathrm{fluct},0}$ -- and their oscillatory parts yielding QOs -- denoted $\tilde{\Delta}$, $\tilde\Omega_\mathrm{MF}(B)$, and $\tilde\Omega_\mathrm{fluct}(B)$ -- which are our primary interest. 
We shall focus on the behavior of oscillations at the fundamental frequency, which is related to the area in reciprocal space in which the unhybridized bands overlap, set by the condition that $\delta[\epsilon_0/(2\omega_c)] =1$, {\it i.e.} a frequency in $1/B$ of $m\epsilon_0/(2e)$.

For the mean-field theory the fundamental frequency oscillation of $\tilde\Omega_\mathrm{MF}(B)$ in the $T\to0$ limit was evaluated in Ref.~\onlinecite{Allocca2022} and was found to be
\begin{equation} \label{eq:OmegaMFtilde}
    \tilde\Omega_\mathrm{MF}(B) \approx -\frac{2\Delta_0 N_\Phi}{\pi} \cos\left(2\pi \frac{\epsilon_0}{2\omega_c}\right)K_1\left(2\pi \frac{\Delta_0}{\omega_c}\right),
\end{equation}
where $N_\Phi = eBA/h$ is the number of electrons in each filled Landau level, and $K_1$ is the modified Bessel function of the second kind.
For the weak field regime $\omega_c \ll 2\pi\Delta_0$ the asymptotic form $K_1(x\gg1) \sim \sqrt{\frac{\pi}{2x}} e^{-x}$ shows exponential suppression in $x=2\pi\Delta_0/\omega_c$.

For the quantum fluctuation corrections beyond mean-field theory the oscillatory contribution $\tilde\Omega_\mathrm{fluct}(B)$ can be obtained by exact numerical evaluation of the full fluctuation free energy \cref{eq:Omegafluct}. 
We will present these results below. 
However, first it is helpful to derive some simplified results based on specific approximations. 
Because fluctuations provide a small contribution to the total free energy by assumption, the entire quantity $\Omega_\mathrm{fluct}$ must be small compared to $\Omega_\mathrm{MF}$. 
Furthermore, examining the dependence of the polarization on frequency and momentum we see that the trace-log in \cref{eq:Omegafluct} is always negative, so the smallness of $\Omega_\mathrm{fluct}$ cannot be attributed to cancellations from contributions at different $q$ --- $\Omega_\mathrm{fluct}$ must be small because $V\hat{\Pi}_q$ is itself small. 
Therefore, to find the dominant contribution we can expand the logarithm in \cref{eq:Omegafluct} and keep just the first term, proportional to the trace of the polarization.  
The remaining sums can be done exactly and in the $T\to0$ limit we obtain
\begin{equation} \label{eq:OmegafluctResult}
    \Omega_\mathrm{fluct}(B) \approx \eval{\frac{N_\Phi^2 V}{4 A} \sum_{l,l'}\left(1-\frac{\xi_l\xi_{l'}}{E_l E_{l'}}\right)}_{\Delta=\Delta_0}.
\end{equation}
All oscillations arise from the second term, which can be evaluated with the Poisson summation formula to show that the oscillation at the fundamental frequency is
\begin{equation} \label{eq:Omegaflucttilde}
    \tilde\Omega_\mathrm{fluct}(B) \approx -2 \rho_F V \Delta_0 \,\delta n_e A \sin\left(2\pi \frac{\epsilon_0}{2\omega_c}\right) K_1\left(2\pi \frac{\Delta_0}{\omega_c}\right),
\end{equation}
where
\begin{multline}
    \delta n_e = \rho_F\left(\sqrt{\left(\Lambda-\frac{\epsilon_0}{2}\right)^2+\Delta_0^2}-\sqrt{\left(\frac{\epsilon_0}{2}\right)^2+\Delta_0^2}\right) \\
    \approx \rho_F(\Lambda-\epsilon_0) = n_v - n_c
\end{multline}
is the difference of the original valence and conduction band electron densities. 
(The details of this calculation and others obtaining the same result are given in the Supplement~\cite{Supplement}.)

We verify the above approximate calculation by comparison with the numerical evaluation of the entire fluctuation free energy \cref{eq:Omegafluct}, from which we extract the oscillatory part~\footnote{Much of this numerical analysis is done using the Julia programming language~\cite{Bezanson2017}.}. 
In our numerical analysis we use $\epsilon_0$ as our unit of energy and set $\Lambda = 10$, $\Delta_0 = 0.06, 0.08,$ or $0.10$ (with $\rho_F V$ then fixed by the zero-field gap equation), and temperature $T=\Delta_0/2$~\footnote{We find that the results for different $T\leq\Delta_0/2$ are nearly indistinguishable, so these numerical results are a very good reflection of the $T=0$ behavior of the system.}.
The numerical and analytic results for the oscillatory parts of the free energy using these parameters are plotted in \cref{fig:oscillations}. 
The close agreement we find between them validates the approximations used to derive \cref{eq:Omegaflucttilde} within this parameter regime.

Comparing the results for $\tilde\Omega_\mathrm{MF}$ and $\tilde\Omega_\mathrm{fluct}$ in \cref{eq:OmegaMFtilde,eq:Omegaflucttilde} and \cref{fig:oscillations}, we find that for weak fields $\omega_c \ll 2\pi\Delta_0$ the oscillatory part of the total free energy of the system can easily be dominated by contributions from quantum fluctuations of the gap.
Both $\tilde\Omega_\mathrm{MF}$ and $\tilde\Omega_\mathrm{fluct}$ are exponentially suppressed for weak fields by the same Bessel function factor, but while the prefactor of $\tilde\Omega_\mathrm{MF}$ has only a linear dependence on the (small) magnetic field strength, $\tilde\Omega_\mathrm{fluct}$ depends on the interaction strength and the imbalance of electron densities between the two bands $\delta n_e$.
This $\delta n_e$ may be very large depending on $\epsilon_0$, setting the carrier density in the system, and $\Lambda$, parameterizing the total valence bandwidth or the total electron density in the filled valence band.

We expect that the effect is very generic. 
It arises from the contributions of electron-electron interactions to the energy offsets of the relevant bands. 
Indeed, the fluctuation contribution \cref{eq:OmegafluctResult} can also be computed as the interaction energy from the original interaction Hamiltonian between conduction and valence electrons by taking expectation values of the respective electron densities in the mean field state,
$E_\mathrm{int} = V\expval{\hat{n}_c}_\mathrm{MF} \expval{\hat{n}_v}_\mathrm{MF}$.
This interaction energy encodes the effect that the band energy of a $c$ electron depends on the occupation of the $v$ electrons, {\it i.e.} the interaction-induced shift of the band edges.  
In the excitonic-insulator state, the bands are hybridized such that each is partially occupied.  
The application of a magnetic field leads to Landau quantization of the occupied states. 
As the field is swept, at fixed total electron number, there is an oscillation in the number difference between the two bands, which, through the differences between inter- and intra-band interactions, leads to oscillatory band energy offsets and hence an oscillation in the total energy. 
The dependence on $V$ shows that the size of these quantum oscillations may be used to probe the size of interactions in these insulating materials. 
Here the signature is in the fundamental oscillation frequency, so differs from theories of interaction-induced harmonics for metallic systems~\cite{Allocca2021,Leeb2023}. 

Perhaps most surprisingly we find that there exists a parameter regime where the amplitude of $\tilde\Omega_\mathrm{fluct}$ can be even larger than the oscillations found for the corresponding gapless system obtained as the $\Delta\to0$ limit of the mean field theory \cref{eq:OmegaMFtilde} that applies when setting the interaction $V=0$ from the start,
\begin{equation} \label{eq:OmegaV=0}
    \tilde\Omega_{V=0}(B) \approx -\frac{N_\Phi \omega_c}{\pi^2} \cos\left(2\pi\frac{\epsilon_0}{2\omega_c}\right),
\end{equation}
which is also shown in \cref{fig:oscillations}.
The oscillatory contributions for the insulator are still exponentially suppressed as $B\to 0$, however for low-electron density materials the regime where the oscillations remain large is in readily accessible ranges of magnetic fields.

To illustrate that the parameter regimes we study here are appropriate for the excitonic insulators that can currently be realized experimentally, we consider the MoSe${}_2$/WSe${}_2$ devices examined in Ref.~\onlinecite{Wang2019}. 
Carrier densities of $\sim10^{12}\,\mathrm{cm}^{-2}$ can be achieved in these TMD double layer systems through gating, and using the effective mass of electrons and holes in their respective bands, $m_{c,v}\approx m_e/2$ with $m_e$ the bare electron mass, this corresponds to $\epsilon_0\sim 20\,\mathrm{meV}$. 
Using these values the range of $\epsilon_0/\omega_c$ plotted in \cref{fig:oscillations} thus corresponds to $B \sim 5-20\,\mathrm{T}$. 
The bandwidth of the WSe${}_2$ valence band, related to $\Lambda$, is $\sim 1\,\mathrm{eV}$~\cite{Le2015}.
The exciton binding energy in these devices, corresponding to our $\Delta_0$, is $\sim100\,\mathrm{meV}$.
This is further into the strong coupling regime than the theory we consider since their goal of room-temperature condensation is dependent on large binding energy, but this is not a fundamental issue.
In principle weaker interactions, putting the system into the BCS regime of our calculations, can be achieved with larger spacing between TMD layers.

We have shown how the nature of quantum oscillations in excitonic insulators is principally determined by quantum fluctuations of the gap. 
Not only are these QOs significantly larger than what is obtained from just a mean field treatment of these systems, for low-carrier-density semiconductors they can be even larger than the oscillations obtained from the non-interacting gapless state from which the excitonic insulator state arises. 
We suggest that the sort of TMD systems already shown to host excitonic insulating states are prime candidates to see this effect realized. 

\begin{acknowledgments}
The authors declare no competing interests. 
We acknowledge helpful conversations with Justin Wilson, Zachary Raines, and Mike Payne, and thank Justin Wilson for helping to speed up our numerical methods.
This work is supported by EPSRC Grant No. EP/P034616/1 and by a Simons Investigator Award.
\end{acknowledgments}

\bibliography{references}

%%%%% Uncomment for arxiv submission
%%%%%% vvvvvvvvvvvvvvv

\input{supplement.tex}

%%%%%% ^^^^^^^^^^^^^^^
%%%%% Uncomment for arxiv submission

\end{document}

%% file: supplement.tex
% \documentclass[aps,prx,reprint,superscriptaddress,longbibliography]{revtex4-2}

% %%%%%USER PACKAGES
% \usepackage{amsmath}
% \usepackage{amssymb}
% \usepackage{bm}
% \usepackage{physics}

% \usepackage{hyperref}
% \usepackage[capitalize]{cleveref}

% \usepackage[usenames]{xcolor}
% \usepackage{graphicx}

% %%%%%USER COMMANDS
% \DeclareMathOperator{\sgn}{sgn}
% \DeclareMathOperator{\arsinh}{arsinh}
% \newcommand{\overbar}[1]{\mkern 1.5mu\overline{\mkern-1.5mu#1\mkern-1.5mu}\mkern 1.5mu}

%\bibliographystyle{apsrev4-2}

%\begin{document}
% \title{Supplement to Fluctuation-Dominated Quantum Oscillations  in Excitonic Insulators}

% \author{Andrew A. Allocca} \email{aallocca@lsu.edu}
% \affiliation{T.C.M. Group, Cavendish Laboratory, University of Cambridge, JJ Thomson Avenue, Cambridge, CB3 0HE, U.K.\looseness=-1}
% \affiliation{Department of Physics and Astronomy, Louisiana State University, Baton Rouge, LA 70803, USA}
% \affiliation{Center for Computation and Technology, Louisiana State University, Baton Rouge, LA 70803, USA}
% \author{Nigel R. Cooper}
% \affiliation{T.C.M. Group, Cavendish Laboratory, University of Cambridge, JJ Thomson Avenue, Cambridge, CB3 0HE, U.K.\looseness=-1}
% \affiliation{Department of Physics and Astronomy, University of Florence, Via G. Sansone 1, 50019 Sesto Fiorentino, Italy\looseness=-1}

% \date{\today}
% \maketitle
%%%%%% ^^^^^^^^^^^^^^^^^^^^^^^^^^^^^^^
%%%%% Comment out for arxiv submission

%%%%% Uncomment for arxiv submission
%%%%%% VVVVVVVVVVVVVVV

\clearpage
\onecolumngrid
\appendix*

%%%%%% ^^^^^^^^^^^^^^^^^^^^^^^^^^^^^^^
%%%%% Uncomment for arxiv submission

\section{Coupling to B field}
To include an external magnetic field $B$ perpendicular to the system we consider minimally coupling the fermionic theory to a static vector potential in the Landau gauge. 
The effect is to quantize the electrons into Landau levels with energies $\xi_l = \omega_c (l + 1/2) -\epsilon_0/2$, where $l = 0,1,2,\dots$ and $\omega_c = eB/m$ is the cyclotron energy, and corresponding wave functions
\begin{equation}
    \Phi_{l,k_y}(x,y) = e^{i k_y y} \phi_l(x-\ell_B^2 k_y),
\end{equation}
using
\begin{equation} 
    \phi_l(x) = \frac{1}{\sqrt{2^l l!}}\left(\frac{1}{\pi \ell_B^2}\right)^{1/4} e^{-x^2/(2 \ell_B^2)} H_l\left(x/\ell_B\right),
\end{equation}
where $H_l(x)$ are the Hermite polynomials and $\ell_B = 1/\sqrt{eB}$ is the magnetic length. 
The expressions obtained above for the system at $B=0$ translate to their nonzero field equivalents with simple substitutions: $\xi_k\to\xi_l$, so the mean field bands are $E_k\to E_l = \sqrt{\xi_l^2+\Delta^2}$, and with our choice of gauge $\sum_\mathbf{k} \to \sum_l\sum_{k_y}$. 
Dependence on $k_x$ is thus replaced by $l$ and the momentum $k_y$ now labels the degenerate states in each Landau level so that $\sum_{k_y} = A B/\Phi_0 \equiv N_\Phi$ is the degeneracy of each Landau level, where $A$ is the system's area and $\Phi_0=h/e$ is the magnetic flux quantum.
In this basis we now use subscript $k$ on fermions for Matsubara frequency and the remaining momentum index, with the Landau level index written separately.

The fluctuation field $\eta_q$, being a neutral bosonic degree of freedom, does not directly couple to an electromagnetic field at the level of minimal coupling, and the primary effect of this change to the basis of Landau levels is to introduce a nontrivial coupling between fluctuations and fermions:
\begin{equation} \label{eq:Seta}
    S_{\eta} = \frac{\beta A}{V}\sum_{q} \bar{\eta}_q \eta_q - \sum_{l,l'}\sum_{q,k} \bar{\Psi}_{l,k+\tfrac{q}{2}} g^{ll'}_{\mathbf{q},k_y}\mqty(0 & \eta_q \\ \bar{\eta}_{-q} & 0) \Psi_{l',k-\tfrac{q}{2}},
\end{equation}
with the coupling
\begin{equation}
    g^{ll'}_{\mathbf{q},k_y} \equiv e^{i \ell_B^2 q_x k_y}\int \dd x\,  \phi_l(x-\ell_B^2\tfrac{q_y}{2}) \phi_{l'}(x+\ell_B^2\tfrac{q_y}{2}) e^{i q_x x}.
\end{equation}
%Note that because the fermionic fields appearing in \cref{eq:Seta0} have different momenta before turning on the field, after doing so they have different Landau level indices. 

\section{Free energies}
Integrating out the fermionic fields for $S_\mathrm{MF}$ yields the mean field free energy, which for $T\to0$ is equal to the energy of the mean field itself plus the sum over all occupied electronic states,
\begin{equation} \label{eq:OmegaMF}
    \Omega_\mathrm{MF}(B) = \frac{\Delta^2A}{V} - N_\Phi \sum_l E_l,
\end{equation}
and the stationary condition on this free energy determining the mean field gap is
\begin{equation} \label{eq:gapEq}
    \frac{1}{V} = \frac{N_\Phi}{A} \sum_{l} \frac{1}{2\sqrt{\xi_l^2 + \Delta^2}}.
\end{equation}
For this equation to have a solution the gap itself must depend on $B$, so that $\Delta = \Delta(B)$. 
The same procedure for $B=0$ would instead produce
\begin{equation} \label{eq:gapEq0}
    \frac{1}{V} = \frac{1}{A}\sum_\mathbf{k} \frac{1}{2\sqrt{\xi_k^2+\Delta_0^2}} = \rho_F \left[\arsinh\left(\frac{\Lambda-\frac{\epsilon_0}{2}}{\Delta_0}\right) +\arsinh\left(\frac{\epsilon_0}{2\Delta_0}\right)\right],
\end{equation}
where we define $\Delta_0$ as the value of the gap at $B=0$. 

We now introduce some general notation and assumptions we will use throughout the rest of our analysis: with the zero-field gap satisfying \cref{eq:gapEq0}, we can define $\delta{\Delta}(B) \equiv \Delta(B)-\Delta_0$ which contains all of the gap's magnetic field dependence.
We assume $\delta\Delta(B)$ vanishes continuously as $B\to0$ and restrict our focus to the regime of magnetic field for which $\abs{\delta\Delta(B)}\ll\Delta_0$.
We denote the specifically oscillatory part of $\delta\Delta(B)$ as $\tilde\Delta(B)$.
For a generic $f(B)$ the two quantities $\delta f(B)$ and $\tilde f(B)$ defined in this way need not be the same, but with the approximations we make in our model we find that the two are equivalent for all quantities we will consider. 
In Ref.~\onlinecite{Allocca2022} this field-dependent component of the gap and its effect on QO in the mean field approximation was explored in detail for model excitonic and Kondo insulators. 

\subsection{Fluctuation free energy}
Turning now to $S_{\eta}$, after integrating out fermions we expand up to second order in the fluctuations and obtain a Gaussian action,
\begin{equation} 
    S_\mathrm{fluct} = \beta A\sum_q (h_{-q}\; \varphi_{-q}) \left(\frac{\hat{\mathbf{1}}}{V} + \hat{\Pi}_q\right) \mqty(h_q \\ \varphi_q),
\end{equation}
where we have put $\eta_q = h_q + i \varphi_q$, with $h_q$ and $\varphi_q$ real bosonic fields representing the Higgs and phase modes respectively. 
The polarization $\hat{\Pi}_q$ now contains all information about coupling to the underlying electrons, and has the form
\begin{multline} \label{eq:polarization}
    \hat{\Pi}_q = \frac{N_\Phi}{2\beta A}\sum_{\epsilon_n}\sum_{l,l'}\frac{\abs{\mel{l}{e^{iq\hat{x}}}{l'}}^2}{[(i\epsilon_n)^2 - E_l^2][(i\epsilon^+_n)^2 - E_{l'}^2]}\\
    \times\mqty((i\epsilon_n+\xi_l)(i\epsilon^+_n-\xi_{l'}) + (i\epsilon_n-\xi_l)(i\epsilon^+_n+\xi_{l'}) + 2\Delta^2 & i\left[(i\epsilon_n+\xi_l)(i\epsilon^+_n-\xi_{l'}) - (i\epsilon_n-\xi_l)(i\epsilon^+_n+\xi_{l'})\right] \\ 
    -i\left[(i\epsilon_n+\xi_l)(i\epsilon^+_n-\xi_{l'}) - (i\epsilon_n-\xi_l)(i\epsilon^+_n+\xi_{l'})\right] & (i\epsilon_n+\xi_l)(i\epsilon^+_n-\xi_{l'}) + (i\epsilon_n-\xi_l)(i\epsilon^+_n+\xi_{l'}) - 2\Delta^2),
\end{multline}
where $\epsilon^+_n = \epsilon_n + \omega_m$ and the squared matrix element is from two factors of the electron-fluctuation coupling $g$,
\begin{equation}
    \frac{1}{2\pi}\sum_{k_y} \,g^{ll'}_{\mathbf{q},k_y} g^{l'l}_{-\mathbf{q},k_y} = N_\Phi \abs{\int \dd x \phi_l(x) \phi_{l'}(x) e^{iqx}}^2 \equiv N_\Phi\abs{\mel{l}{e^{iq\hat{x}}}{l'}}^2.
\end{equation}
This polarization has a similar form to what has been obtained previously when analyzing the fluctuations in BCS theory, but there are some notable differences. 
In particular, particle-hole symmetry ensures that the Higgs and phase modes decouple in systems at $B=0$, but here there are nontrivial off-diagonal elements in $\hat{\Pi}_q$ due to the breaking of time reversal symmetry by the magnetic field and these modes are mixed in general. 
To obtain the fluctuation contribution to the free energy we finally integrate out the bosonic fields and find
\begin{equation}
    \Omega_\mathrm{fluct}(B) = \frac{1}{2\beta}\sum_q \tr\ln\left(\hat{\mathbf{1}}+V\hat{\Pi}_q\right).
\end{equation} 
We take the $T\to0$ limit of this quantity when evaluating it explicitly.

\subsection{Oscillatory free energies}
We are specifically interested in the oscillatory part of the free energy, $\tilde\Omega(B,\Delta(B))$, which is responsible for quantum oscillations of thermodynamic quantities like the magnetization via $\tilde{M}(B) = -\partial\tilde{\Omega}(B)/\partial B$. 
We can isolate this part as in Ref.~\onlinecite{Allocca2022} by first using $\Delta(B) = \Delta_0 + \tilde\Delta(B)$ with $\abs*{\tilde\Delta(B)} \ll \Delta_0$ to expand the free energy around $\Delta=\Delta_0$ in powers of $\tilde\Delta(B)$, then separating $\Omega(B,\Delta_0)$ into its $B=0$ part $\Omega_0(\Delta_0)$ and its $B$-dependent oscillatory part $\tilde\Omega(B,\Delta_0)$.
As for $\Delta_0$ and $\tilde\Delta$, we assume that $\abs*{\tilde{\Omega}}\ll \Omega_0$, which we verify \emph{post hoc} by numerically evaluating the free energy without approximation. 
Also separating the mean field and fluctuation parts of the free energy, altogether we obtain
\begin{equation} \label{eq:OmegafluctSUPP}
    \Omega(B,\Delta(B)) \approx \Omega_{\mathrm{MF},0} + \Omega_{\mathrm{fluct},0} + \tilde{\Omega}_\mathrm{MF}(B) + \tilde{\Omega}_\mathrm{fluct}(B) + \tilde\Delta(B)\,\frac{\partial}{\partial\Delta_0}\left(\Omega_{\mathrm{MF},0} + \Omega_{\mathrm{fluct},0}\right),
\end{equation}
where every $\Omega$ is evaluated at $\Delta = \Delta_0$. 
Any further terms in this expansion are necessarily at least second order in small oscillatory quantities, which contribute only to second and higher harmonic oscillations.
Since our interest is in oscillations at the fundamental frequency, we drop these terms. 
The final term this expression, proportional to $\tilde\Delta$, also does not contribute further; $\Delta_0$ is the value for which the $B=0$ part of the free energy is stationary, so the derivative vanishes by definition. 
For the sort of model we have here it can be shown that the correction to the gap from fluctuations above the purely mean field value is negligible~\cite{Kos2004,Hoyer2018}, so to a very good approximation $\Delta_0$ is determined from just the mean field term, recovering exactly \cref{eq:gapEq}.
Thus, the largest oscillatory part of the free energy is the sum of two terms, $\tilde\Omega_\mathrm{MF}(B)$ and $\tilde\Omega_\mathrm{fluct}(B)$, which are just the oscillatory parts of the mean field and fluctuation free energies evaluated with the zero-field gap $\Delta_0$. 

As discussed in the main text, the fact that \cref{eq:OmegafluctSUPP} is small can be attributed to $V\hat\Pi_q$ being small (it has eigenvalues with absolute value $\ll1$).
Using this and the above approximations we can expand the log to first order and write
\begin{align*}
    \Omega_\mathrm{fluct}(B) &\approx \frac{V}{2\beta}\sum_q \eval{\tr\hat\Pi_q}_{\Delta=\Delta_0} \\
    &= \frac{N_\Phi V}{2\beta^2A}\sum_q \sum_{\epsilon_n}\sum_{l,l'} \eval{\frac{\abs{\mel{l}{e^{iq\hat{x}}}{l'}}^2}{[(i\epsilon_n)^2 - E_l^2][(i\epsilon^+_n)^2 - E_{l'}^2]} \left[(i\epsilon_n+\xi_l)(i\epsilon^+_n-\xi_{l'}) + (i\epsilon_n-\xi_l)(i\epsilon^+_n + \xi_{l'})\right]}_{\Delta=\Delta_0} \\
    &= \frac{N_\Phi^2 V}{\beta^2A}\sum_{\epsilon_n,\epsilon_{n'}}\sum_{l,l'} \eval{\frac{i\epsilon_n+\xi_l}{(i\epsilon_n)^2 - E_l^2} \frac{i\epsilon_{n'}-\xi_{l'}}{(i\epsilon_{n'})^2 - E_{l'}^2}}_{\Delta=\Delta_0} \\
    &= \frac{N_\Phi^2 V}{4A} \sum_{l,l'} \eval{\left(1-\frac{\xi_l}{E_l}\right)\left(1+\frac{\xi_{l'}}{E_{l'}}\right)}_{\Delta=\Delta_0} \\
    &=\eval{\frac{N_\Phi^2 V}{4A} \sum_{l,l'} \left(1-\frac{\xi_l\xi_{l'}}{E_l E_{l'}}\right)}_{\Delta=\Delta_0}.
\end{align*}
Because of these approximations, the off-diagonal elements of $\hat\Pi_q$ do not contribute, so the coupling between Higgs and phase modes are found not to be relevant for the dominant contribution to thermodynamic QO. 
Going the third line we perform the sum over $\mathbf{q}$, which can be done exactly, and seeing that $\omega_m$ only appears in the combination $\epsilon^+_n = \epsilon_n+\omega_m$ we exchange the sum over $\omega_m$ with a sum over this fermionic frequency redefined as $\epsilon_n'$. 
The Matsubara sums can then be performed exactly, giving the main result. 

\section{Alternative Calculation}
\begin{figure}[!h]
   \centering
   \includegraphics[width=0.5\columnwidth]{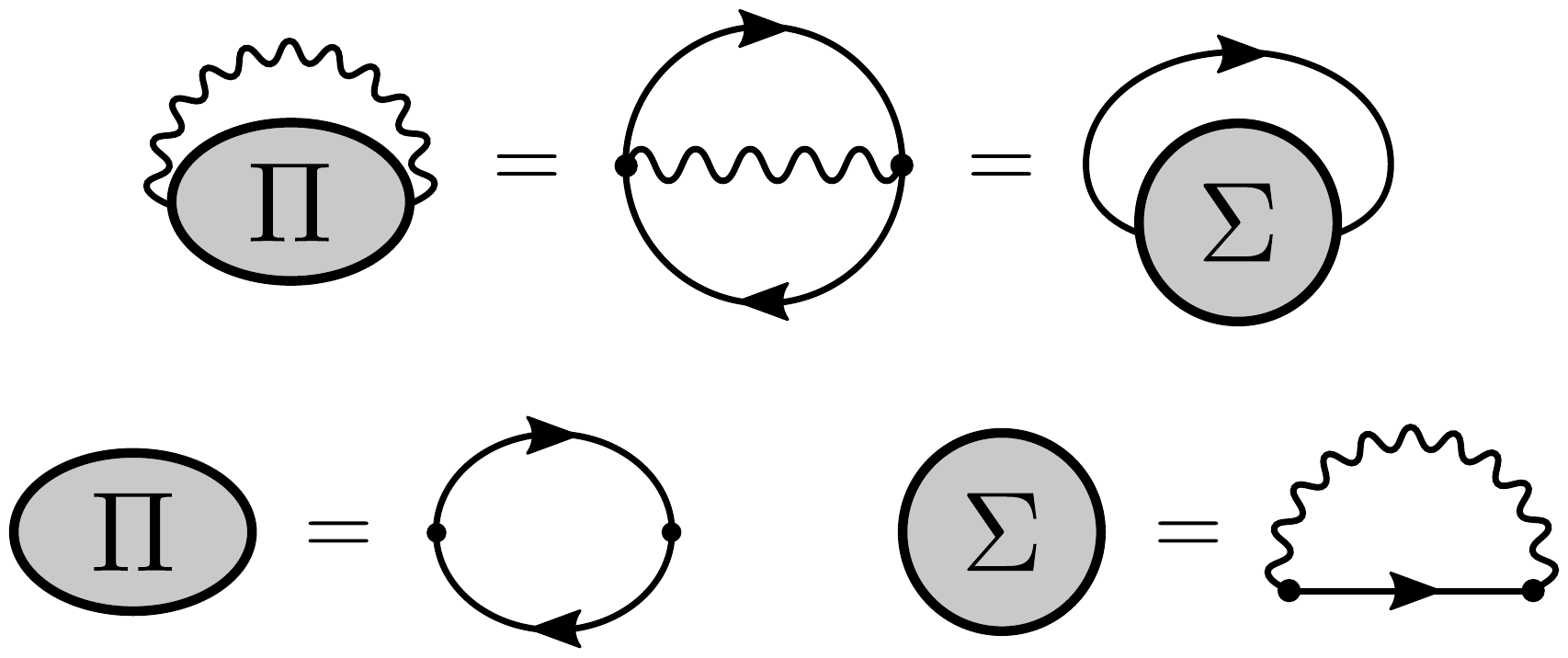}
   \caption{The first-order loop diagram giving the main result for the fluctuation free energy, \cref{eq:OmegafluctResult}, expressed in terms of both the polarization $\Pi$ of the bosonic fluctuations and the self-energy $\Sigma$ of the electrons.
   The straight solid lines are fermion Green's functions, the wavy lines are the interaction parameterized by $V$, and the dots are the electron-fluctuation coupling $g \,\hat\sigma_{x,y}$.}
    \label{fig:diagrams}
\end{figure}

There is a complementary way to approach the calculation of our main result. 
Starting from the mean-field and fluctuation actions appropriately rewritten in the basis of Landau levels, we can include the effect of the fluctuating modes by evaluating the Fock self-energy they provide to electrons in the conduction and valence bands
Mean field theory is equivalent to setting the corresponding Hartree self-energy equal to the gap $\Delta$, and indeed if we evaluate the Hartree self-energy it is equal to $\Delta$ if the gap equation holds.
Then computing the contribution of the self-energy to the free energy at first order in $V$ and putting $\Delta\approx\Delta_0$ we recover exactly the main result,
\begin{equation}
    \Omega_\mathrm{fluct}(B) \approx \eval{\frac{N_\Phi^2 V}{4A} \sum_{l,l'}\left(1-\frac{\xi_l\xi_{l'}}{E_l E_{l'}}\right)}_{\Delta=\Delta_0}.
\end{equation}
This calculation in terms of a self-energy and the calculation discussed in the main text  and this one in terms of a self-energy are precisely the two complementary ways of evaluating the loop diagram shown in \cref{fig:diagrams}.

%\bibliography{supplement_references}
%%%%% Comment out for arxiv submission
%%%%%% VVVVVVVVVVVVVVV

% \bibliography{references}

% \end{document}

%%%%%% ^^^^^^^^^^^^^^^^^^^^^^^^^^^^^^^
%%%%% Comment out for arxiv submission

%% file: main.bbl
\begin{thebibliography}{55}%
\makeatletter
\providecommand \@ifxundefined [1]{%
 \@ifx{#1\undefined}
}%
\providecommand \@ifnum [1]{%
 \ifnum #1\expandafter \@firstoftwo
 \else \expandafter \@secondoftwo
 \fi
}%
\providecommand \@ifx [1]{%
 \ifx #1\expandafter \@firstoftwo
 \else \expandafter \@secondoftwo
 \fi
}%
\providecommand \natexlab [1]{#1}%
\providecommand \enquote  [1]{``#1''}%
\providecommand \bibnamefont  [1]{#1}%
\providecommand \bibfnamefont [1]{#1}%
\providecommand \citenamefont [1]{#1}%
\providecommand \href@noop [0]{\@secondoftwo}%
\providecommand \href [0]{\begingroup \@sanitize@url \@href}%
\providecommand \@href[1]{\@@startlink{#1}\@@href}%
\providecommand \@@href[1]{\endgroup#1\@@endlink}%
\providecommand \@sanitize@url [0]{\catcode `\\12\catcode `\$12\catcode
  `\&12\catcode `\#12\catcode `\^12\catcode `\_12\catcode `\%12\relax}%
\providecommand \@@startlink[1]{}%
\providecommand \@@endlink[0]{}%
\providecommand \url  [0]{\begingroup\@sanitize@url \@url }%
\providecommand \@url [1]{\endgroup\@href {#1}{\urlprefix }}%
\providecommand \urlprefix  [0]{URL }%
\providecommand \Eprint [0]{\href }%
\providecommand \doibase [0]{https://doi.org/}%
\providecommand \selectlanguage [0]{\@gobble}%
\providecommand \bibinfo  [0]{\@secondoftwo}%
\providecommand \bibfield  [0]{\@secondoftwo}%
\providecommand \translation [1]{[#1]}%
\providecommand \BibitemOpen [0]{}%
\providecommand \bibitemStop [0]{}%
\providecommand \bibitemNoStop [0]{.\EOS\space}%
\providecommand \EOS [0]{\spacefactor3000\relax}%
\providecommand \BibitemShut  [1]{\csname bibitem#1\endcsname}%
\let\auto@bib@innerbib\@empty
%</preamble>
\bibitem [{\citenamefont {Cloizeaux}(1965)}]{Cloizeaux1965}%
  \BibitemOpen
  \bibfield  {author} {\bibinfo {author} {\bibfnamefont {J.~D.}\ \bibnamefont
  {Cloizeaux}},\ }\bibfield  {title} {\bibinfo {title} {Exciton instability and
  crystallographic anomalies in semiconductors},\ }\href
  {https://doi.org/https://doi.org/10.1016/0022-3697(65)90153-8} {\bibfield
  {journal} {\bibinfo  {journal} {Journal of Physics and Chemistry of Solids}\
  }\textbf {\bibinfo {volume} {26}},\ \bibinfo {pages} {259} (\bibinfo {year}
  {1965})}\BibitemShut {NoStop}%
\bibitem [{\citenamefont {J\'erome}\ \emph {et~al.}(1967)\citenamefont
  {J\'erome}, \citenamefont {Rice},\ and\ \citenamefont {Kohn}}]{Jerome1967}%
  \BibitemOpen
  \bibfield  {author} {\bibinfo {author} {\bibfnamefont {D.}~\bibnamefont
  {J\'erome}}, \bibinfo {author} {\bibfnamefont {T.~M.}\ \bibnamefont {Rice}},\
  and\ \bibinfo {author} {\bibfnamefont {W.}~\bibnamefont {Kohn}},\ }\bibfield
  {title} {\bibinfo {title} {Excitonic insulator},\ }\href
  {https://doi.org/10.1103/PhysRev.158.462} {\bibfield  {journal} {\bibinfo
  {journal} {Phys. Rev.}\ }\textbf {\bibinfo {volume} {158}},\ \bibinfo {pages}
  {462} (\bibinfo {year} {1967})}\BibitemShut {NoStop}%
\bibitem [{\citenamefont {Keldysh}\ and\ \citenamefont
  {Kozlov}(1968)}]{Keldysh1968}%
  \BibitemOpen
  \bibfield  {author} {\bibinfo {author} {\bibfnamefont {L.}~\bibnamefont
  {Keldysh}}\ and\ \bibinfo {author} {\bibfnamefont {A.}~\bibnamefont
  {Kozlov}},\ }\bibfield  {title} {\bibinfo {title} {Collective properties of
  excitons in semiconductors},\ }\href@noop {} {\bibfield  {journal} {\bibinfo
  {journal} {Sov. Phys. JETP}\ }\textbf {\bibinfo {volume} {27}},\ \bibinfo
  {pages} {521} (\bibinfo {year} {1968})}\BibitemShut {NoStop}%
\bibitem [{\citenamefont {Halperin}\ and\ \citenamefont
  {Rice}(1968)}]{Halperin1968}%
  \BibitemOpen
  \bibfield  {author} {\bibinfo {author} {\bibfnamefont {B.~I.}\ \bibnamefont
  {Halperin}}\ and\ \bibinfo {author} {\bibfnamefont {T.~M.}\ \bibnamefont
  {Rice}},\ }\bibfield  {title} {\bibinfo {title} {Possible anomalies at a
  semimetal-semiconductor transistion},\ }\href
  {https://doi.org/10.1103/RevModPhys.40.755} {\bibfield  {journal} {\bibinfo
  {journal} {Rev. Mod. Phys.}\ }\textbf {\bibinfo {volume} {40}},\ \bibinfo
  {pages} {755} (\bibinfo {year} {1968})}\BibitemShut {NoStop}%
\bibitem [{\citenamefont {Jia}\ \emph {et~al.}(2022)\citenamefont {Jia},
  \citenamefont {Wang}, \citenamefont {Chiu}, \citenamefont {Song},
  \citenamefont {Yu}, \citenamefont {J{\"a}ck}, \citenamefont {Lei},
  \citenamefont {Klemenz}, \citenamefont {Cevallos}, \citenamefont {Onyszczak},
  \citenamefont {Fishchenko}, \citenamefont {Liu}, \citenamefont {Farahi},
  \citenamefont {Xie}, \citenamefont {Xu}, \citenamefont {Watanabe},
  \citenamefont {Taniguchi}, \citenamefont {Bernevig}, \citenamefont {Cava},
  \citenamefont {Schoop}, \citenamefont {Yazdani},\ and\ \citenamefont
  {Wu}}]{Jia2022}%
  \BibitemOpen
  \bibfield  {author} {\bibinfo {author} {\bibfnamefont {Y.}~\bibnamefont
  {Jia}}, \bibinfo {author} {\bibfnamefont {P.}~\bibnamefont {Wang}}, \bibinfo
  {author} {\bibfnamefont {C.-L.}\ \bibnamefont {Chiu}}, \bibinfo {author}
  {\bibfnamefont {Z.}~\bibnamefont {Song}}, \bibinfo {author} {\bibfnamefont
  {G.}~\bibnamefont {Yu}}, \bibinfo {author} {\bibfnamefont {B.}~\bibnamefont
  {J{\"a}ck}}, \bibinfo {author} {\bibfnamefont {S.}~\bibnamefont {Lei}},
  \bibinfo {author} {\bibfnamefont {S.}~\bibnamefont {Klemenz}}, \bibinfo
  {author} {\bibfnamefont {F.~A.}\ \bibnamefont {Cevallos}}, \bibinfo {author}
  {\bibfnamefont {M.}~\bibnamefont {Onyszczak}}, \bibinfo {author}
  {\bibfnamefont {N.}~\bibnamefont {Fishchenko}}, \bibinfo {author}
  {\bibfnamefont {X.}~\bibnamefont {Liu}}, \bibinfo {author} {\bibfnamefont
  {G.}~\bibnamefont {Farahi}}, \bibinfo {author} {\bibfnamefont
  {F.}~\bibnamefont {Xie}}, \bibinfo {author} {\bibfnamefont {Y.}~\bibnamefont
  {Xu}}, \bibinfo {author} {\bibfnamefont {K.}~\bibnamefont {Watanabe}},
  \bibinfo {author} {\bibfnamefont {T.}~\bibnamefont {Taniguchi}}, \bibinfo
  {author} {\bibfnamefont {B.~A.}\ \bibnamefont {Bernevig}}, \bibinfo {author}
  {\bibfnamefont {R.~J.}\ \bibnamefont {Cava}}, \bibinfo {author}
  {\bibfnamefont {L.~M.}\ \bibnamefont {Schoop}}, \bibinfo {author}
  {\bibfnamefont {A.}~\bibnamefont {Yazdani}},\ and\ \bibinfo {author}
  {\bibfnamefont {S.}~\bibnamefont {Wu}},\ }\bibfield  {title} {\bibinfo
  {title} {Evidence for a monolayer excitonic insulator},\ }\href
  {https://doi.org/10.1038/s41567-021-01422-w} {\bibfield  {journal} {\bibinfo
  {journal} {Nature Physics}\ }\textbf {\bibinfo {volume} {18}},\ \bibinfo
  {pages} {87} (\bibinfo {year} {2022})}\BibitemShut {NoStop}%
\bibitem [{\citenamefont {Wang}\ \emph {et~al.}(2019)\citenamefont {Wang},
  \citenamefont {Rhodes}, \citenamefont {Watanabe}, \citenamefont {Taniguchi},
  \citenamefont {Hone}, \citenamefont {Shan},\ and\ \citenamefont
  {Mak}}]{Wang2019}%
  \BibitemOpen
  \bibfield  {author} {\bibinfo {author} {\bibfnamefont {Z.}~\bibnamefont
  {Wang}}, \bibinfo {author} {\bibfnamefont {D.~A.}\ \bibnamefont {Rhodes}},
  \bibinfo {author} {\bibfnamefont {K.}~\bibnamefont {Watanabe}}, \bibinfo
  {author} {\bibfnamefont {T.}~\bibnamefont {Taniguchi}}, \bibinfo {author}
  {\bibfnamefont {J.~C.}\ \bibnamefont {Hone}}, \bibinfo {author}
  {\bibfnamefont {J.}~\bibnamefont {Shan}},\ and\ \bibinfo {author}
  {\bibfnamefont {K.~F.}\ \bibnamefont {Mak}},\ }\bibfield  {title} {\bibinfo
  {title} {Evidence of high-temperature exciton condensation in two-dimensional
  atomic double layers},\ }\href {https://doi.org/10.1038/s41586-019-1591-7}
  {\bibfield  {journal} {\bibinfo  {journal} {Nature}\ }\textbf {\bibinfo
  {volume} {574}},\ \bibinfo {pages} {76} (\bibinfo {year} {2019})}\BibitemShut
  {NoStop}%
\bibitem [{\citenamefont {Ma}\ \emph {et~al.}(2021)\citenamefont {Ma},
  \citenamefont {Nguyen}, \citenamefont {Wang}, \citenamefont {Zeng},
  \citenamefont {Watanabe}, \citenamefont {Taniguchi}, \citenamefont
  {MacDonald}, \citenamefont {Mak},\ and\ \citenamefont {Shan}}]{Ma2021}%
  \BibitemOpen
  \bibfield  {author} {\bibinfo {author} {\bibfnamefont {L.}~\bibnamefont
  {Ma}}, \bibinfo {author} {\bibfnamefont {P.~X.}\ \bibnamefont {Nguyen}},
  \bibinfo {author} {\bibfnamefont {Z.}~\bibnamefont {Wang}}, \bibinfo {author}
  {\bibfnamefont {Y.}~\bibnamefont {Zeng}}, \bibinfo {author} {\bibfnamefont
  {K.}~\bibnamefont {Watanabe}}, \bibinfo {author} {\bibfnamefont
  {T.}~\bibnamefont {Taniguchi}}, \bibinfo {author} {\bibfnamefont {A.~H.}\
  \bibnamefont {MacDonald}}, \bibinfo {author} {\bibfnamefont {K.~F.}\
  \bibnamefont {Mak}},\ and\ \bibinfo {author} {\bibfnamefont {J.}~\bibnamefont
  {Shan}},\ }\bibfield  {title} {\bibinfo {title} {Strongly correlated
  excitonic insulator in atomic double layers},\ }\href
  {https://doi.org/10.1038/s41586-021-03947-9} {\bibfield  {journal} {\bibinfo
  {journal} {Nature}\ }\textbf {\bibinfo {volume} {598}},\ \bibinfo {pages}
  {585} (\bibinfo {year} {2021})}\BibitemShut {NoStop}%
\bibitem [{\citenamefont {Manzeli}\ \emph {et~al.}(2017)\citenamefont
  {Manzeli}, \citenamefont {Ovchinnikov}, \citenamefont {Pasquier},
  \citenamefont {Yazyev},\ and\ \citenamefont {Kis}}]{Manzeli2017}%
  \BibitemOpen
  \bibfield  {author} {\bibinfo {author} {\bibfnamefont {S.}~\bibnamefont
  {Manzeli}}, \bibinfo {author} {\bibfnamefont {D.}~\bibnamefont
  {Ovchinnikov}}, \bibinfo {author} {\bibfnamefont {D.}~\bibnamefont
  {Pasquier}}, \bibinfo {author} {\bibfnamefont {O.~V.}\ \bibnamefont
  {Yazyev}},\ and\ \bibinfo {author} {\bibfnamefont {A.}~\bibnamefont {Kis}},\
  }\bibfield  {title} {\bibinfo {title} {2d transition metal dichalcogenides},\
  }\href {https://doi.org/10.1038/natrevmats.2017.33} {\bibfield  {journal}
  {\bibinfo  {journal} {Nature Reviews Materials}\ }\textbf {\bibinfo {volume}
  {2}},\ \bibinfo {pages} {17033} (\bibinfo {year} {2017})}\BibitemShut
  {NoStop}%
\bibitem [{\citenamefont {Menth}\ \emph {et~al.}(1969)\citenamefont {Menth},
  \citenamefont {Buehler},\ and\ \citenamefont {Geballe}}]{Menth1969}%
  \BibitemOpen
  \bibfield  {author} {\bibinfo {author} {\bibfnamefont {A.}~\bibnamefont
  {Menth}}, \bibinfo {author} {\bibfnamefont {E.}~\bibnamefont {Buehler}},\
  and\ \bibinfo {author} {\bibfnamefont {T.~H.}\ \bibnamefont {Geballe}},\
  }\bibfield  {title} {\bibinfo {title} {{Magnetic and Semiconducting
  Properties of $\mathrm{SmB}_{6}$}},\ }\href
  {https://doi.org/10.1103/PhysRevLett.22.295} {\bibfield  {journal} {\bibinfo
  {journal} {Phys. Rev. Lett.}\ }\textbf {\bibinfo {volume} {22}},\ \bibinfo
  {pages} {295} (\bibinfo {year} {1969})}\BibitemShut {NoStop}%
\bibitem [{\citenamefont {Hewson}(1993)}]{Hewson1993}%
  \BibitemOpen
  \bibfield  {author} {\bibinfo {author} {\bibfnamefont {A.~C.}\ \bibnamefont
  {Hewson}},\ }\href {https://doi.org/10.1017/CBO9780511470752} {\emph
  {\bibinfo {title} {The Kondo Problem to Heavy Fermions}}},\ Cambridge Studies
  in Magnetism\ (\bibinfo  {publisher} {Cambridge University Press},\ \bibinfo
  {year} {1993})\BibitemShut {NoStop}%
\bibitem [{\citenamefont {Dzero}\ \emph {et~al.}(2010)\citenamefont {Dzero},
  \citenamefont {Sun}, \citenamefont {Galitski},\ and\ \citenamefont
  {Coleman}}]{Dzero2010}%
  \BibitemOpen
  \bibfield  {author} {\bibinfo {author} {\bibfnamefont {M.}~\bibnamefont
  {Dzero}}, \bibinfo {author} {\bibfnamefont {K.}~\bibnamefont {Sun}}, \bibinfo
  {author} {\bibfnamefont {V.}~\bibnamefont {Galitski}},\ and\ \bibinfo
  {author} {\bibfnamefont {P.}~\bibnamefont {Coleman}},\ }\bibfield  {title}
  {\bibinfo {title} {{Topological Kondo Insulators}},\ }\href
  {https://doi.org/10.1103/PhysRevLett.104.106408} {\bibfield  {journal}
  {\bibinfo  {journal} {Phys. Rev. Lett.}\ }\textbf {\bibinfo {volume} {104}},\
  \bibinfo {pages} {106408} (\bibinfo {year} {2010})}\BibitemShut {NoStop}%
\bibitem [{\citenamefont {Dzero}\ \emph {et~al.}(2016)\citenamefont {Dzero},
  \citenamefont {Xia}, \citenamefont {Galitski},\ and\ \citenamefont
  {Coleman}}]{Dzero2016}%
  \BibitemOpen
  \bibfield  {author} {\bibinfo {author} {\bibfnamefont {M.}~\bibnamefont
  {Dzero}}, \bibinfo {author} {\bibfnamefont {J.}~\bibnamefont {Xia}}, \bibinfo
  {author} {\bibfnamefont {V.}~\bibnamefont {Galitski}},\ and\ \bibinfo
  {author} {\bibfnamefont {P.}~\bibnamefont {Coleman}},\ }\bibfield  {title}
  {\bibinfo {title} {{Topological Kondo Insulators}},\ }\href
  {https://doi.org/10.1146/annurev-conmatphys-031214-014749} {\bibfield
  {journal} {\bibinfo  {journal} {Annual Review of Condensed Matter Physics}\
  }\textbf {\bibinfo {volume} {7}},\ \bibinfo {pages} {249} (\bibinfo {year}
  {2016})}\BibitemShut {NoStop}%
\bibitem [{\citenamefont {de~Haas}\ and\ \citenamefont {van
  Alphen}(1930)}]{deHaas1930}%
  \BibitemOpen
  \bibfield  {author} {\bibinfo {author} {\bibfnamefont {W.~J.}\ \bibnamefont
  {de~Haas}}\ and\ \bibinfo {author} {\bibfnamefont {P.~M.}\ \bibnamefont {van
  Alphen}},\ }\bibfield  {title} {\bibinfo {title} {{The dependence of the
  susceptibility of diamagnetic metals upon the field}},\ }\href@noop {}
  {\bibfield  {journal} {\bibinfo  {journal} {Proc. Neth. R. Acad. Sci.}\
  }\textbf {\bibinfo {volume} {33}},\ \bibinfo {pages} {1106} (\bibinfo {year}
  {1930})}\BibitemShut {NoStop}%
\bibitem [{\citenamefont {Li}\ \emph {et~al.}(2014)\citenamefont {Li},
  \citenamefont {Xiang}, \citenamefont {Yu}, \citenamefont {Asaba},
  \citenamefont {Lawson}, \citenamefont {Cai}, \citenamefont {Tinsman},
  \citenamefont {Berkley}, \citenamefont {Wolgast}, \citenamefont {Eo},
  \citenamefont {Kim}, \citenamefont {Kurdak}, \citenamefont {Allen},
  \citenamefont {Sun}, \citenamefont {Chen}, \citenamefont {Wang},
  \citenamefont {Fisk},\ and\ \citenamefont {Li}}]{Li2014}%
  \BibitemOpen
  \bibfield  {author} {\bibinfo {author} {\bibfnamefont {G.}~\bibnamefont
  {Li}}, \bibinfo {author} {\bibfnamefont {Z.}~\bibnamefont {Xiang}}, \bibinfo
  {author} {\bibfnamefont {F.}~\bibnamefont {Yu}}, \bibinfo {author}
  {\bibfnamefont {T.}~\bibnamefont {Asaba}}, \bibinfo {author} {\bibfnamefont
  {B.}~\bibnamefont {Lawson}}, \bibinfo {author} {\bibfnamefont
  {P.}~\bibnamefont {Cai}}, \bibinfo {author} {\bibfnamefont {C.}~\bibnamefont
  {Tinsman}}, \bibinfo {author} {\bibfnamefont {A.}~\bibnamefont {Berkley}},
  \bibinfo {author} {\bibfnamefont {S.}~\bibnamefont {Wolgast}}, \bibinfo
  {author} {\bibfnamefont {Y.~S.}\ \bibnamefont {Eo}}, \bibinfo {author}
  {\bibfnamefont {D.-J.}\ \bibnamefont {Kim}}, \bibinfo {author} {\bibfnamefont
  {C.}~\bibnamefont {Kurdak}}, \bibinfo {author} {\bibfnamefont {J.~W.}\
  \bibnamefont {Allen}}, \bibinfo {author} {\bibfnamefont {K.}~\bibnamefont
  {Sun}}, \bibinfo {author} {\bibfnamefont {X.~H.}\ \bibnamefont {Chen}},
  \bibinfo {author} {\bibfnamefont {Y.~Y.}\ \bibnamefont {Wang}}, \bibinfo
  {author} {\bibfnamefont {Z.}~\bibnamefont {Fisk}},\ and\ \bibinfo {author}
  {\bibfnamefont {L.}~\bibnamefont {Li}},\ }\bibfield  {title} {\bibinfo
  {title} {{Two-dimensional Fermi surfaces in Kondo insulator
  $\mathrm{SmB}_6$}},\ }\href {https://doi.org/10.1126/science.1250366}
  {\bibfield  {journal} {\bibinfo  {journal} {Science}\ }\textbf {\bibinfo
  {volume} {346}},\ \bibinfo {pages} {1208} (\bibinfo {year}
  {2014})}\BibitemShut {NoStop}%
\bibitem [{\citenamefont {Tan}\ \emph {et~al.}(2015)\citenamefont {Tan},
  \citenamefont {Hsu}, \citenamefont {Zeng}, \citenamefont {Hatnean},
  \citenamefont {Harrison}, \citenamefont {Zhu}, \citenamefont {Hartstein},
  \citenamefont {Kiourlappou}, \citenamefont {Srivastava}, \citenamefont
  {Johannes}, \citenamefont {Murphy}, \citenamefont {Park}, \citenamefont
  {Balicas}, \citenamefont {Lonzarich}, \citenamefont {Balakrishnan},\ and\
  \citenamefont {Sebastian}}]{Tan2015}%
  \BibitemOpen
  \bibfield  {author} {\bibinfo {author} {\bibfnamefont {B.~S.}\ \bibnamefont
  {Tan}}, \bibinfo {author} {\bibfnamefont {Y.-T.}\ \bibnamefont {Hsu}},
  \bibinfo {author} {\bibfnamefont {B.}~\bibnamefont {Zeng}}, \bibinfo {author}
  {\bibfnamefont {M.~C.}\ \bibnamefont {Hatnean}}, \bibinfo {author}
  {\bibfnamefont {N.}~\bibnamefont {Harrison}}, \bibinfo {author}
  {\bibfnamefont {Z.}~\bibnamefont {Zhu}}, \bibinfo {author} {\bibfnamefont
  {M.}~\bibnamefont {Hartstein}}, \bibinfo {author} {\bibfnamefont
  {M.}~\bibnamefont {Kiourlappou}}, \bibinfo {author} {\bibfnamefont
  {A.}~\bibnamefont {Srivastava}}, \bibinfo {author} {\bibfnamefont {M.~D.}\
  \bibnamefont {Johannes}}, \bibinfo {author} {\bibfnamefont {T.~P.}\
  \bibnamefont {Murphy}}, \bibinfo {author} {\bibfnamefont {J.-H.}\
  \bibnamefont {Park}}, \bibinfo {author} {\bibfnamefont {L.}~\bibnamefont
  {Balicas}}, \bibinfo {author} {\bibfnamefont {G.~G.}\ \bibnamefont
  {Lonzarich}}, \bibinfo {author} {\bibfnamefont {G.}~\bibnamefont
  {Balakrishnan}},\ and\ \bibinfo {author} {\bibfnamefont {S.~E.}\ \bibnamefont
  {Sebastian}},\ }\bibfield  {title} {\bibinfo {title} {{Unconventional Fermi
  surface in an insulating state}},\ }\href
  {https://doi.org/10.1126/science.aaa7974} {\bibfield  {journal} {\bibinfo
  {journal} {Science}\ }\textbf {\bibinfo {volume} {349}},\ \bibinfo {pages}
  {287} (\bibinfo {year} {2015})}\BibitemShut {NoStop}%
\bibitem [{\citenamefont {Hartstein}\ \emph {et~al.}(2018)\citenamefont
  {Hartstein}, \citenamefont {Toews}, \citenamefont {Hsu}, \citenamefont
  {Zeng}, \citenamefont {Chen}, \citenamefont {Hatnean}, \citenamefont {Zhang},
  \citenamefont {Nakamura}, \citenamefont {Padgett}, \citenamefont
  {Rodway-Gant}, \citenamefont {Berk}, \citenamefont {Kingston}, \citenamefont
  {Zhang}, \citenamefont {Chan}, \citenamefont {Yamashita}, \citenamefont
  {Sakakibara}, \citenamefont {Takano}, \citenamefont {Park}, \citenamefont
  {Balicas}, \citenamefont {Harrison}, \citenamefont {Shitsevalova},
  \citenamefont {Balakrishnan}, \citenamefont {Lonzarich}, \citenamefont
  {Hill}, \citenamefont {Sutherland},\ and\ \citenamefont
  {Sebastian}}]{Hartstein2018}%
  \BibitemOpen
  \bibfield  {author} {\bibinfo {author} {\bibfnamefont {M.}~\bibnamefont
  {Hartstein}}, \bibinfo {author} {\bibfnamefont {W.~H.}\ \bibnamefont
  {Toews}}, \bibinfo {author} {\bibfnamefont {Y.-T.}\ \bibnamefont {Hsu}},
  \bibinfo {author} {\bibfnamefont {B.}~\bibnamefont {Zeng}}, \bibinfo {author}
  {\bibfnamefont {X.}~\bibnamefont {Chen}}, \bibinfo {author} {\bibfnamefont
  {M.~C.}\ \bibnamefont {Hatnean}}, \bibinfo {author} {\bibfnamefont {Q.~R.}\
  \bibnamefont {Zhang}}, \bibinfo {author} {\bibfnamefont {S.}~\bibnamefont
  {Nakamura}}, \bibinfo {author} {\bibfnamefont {A.~S.}\ \bibnamefont
  {Padgett}}, \bibinfo {author} {\bibfnamefont {G.}~\bibnamefont
  {Rodway-Gant}}, \bibinfo {author} {\bibfnamefont {J.}~\bibnamefont {Berk}},
  \bibinfo {author} {\bibfnamefont {M.~K.}\ \bibnamefont {Kingston}}, \bibinfo
  {author} {\bibfnamefont {G.~H.}\ \bibnamefont {Zhang}}, \bibinfo {author}
  {\bibfnamefont {M.~K.}\ \bibnamefont {Chan}}, \bibinfo {author}
  {\bibfnamefont {S.}~\bibnamefont {Yamashita}}, \bibinfo {author}
  {\bibfnamefont {T.}~\bibnamefont {Sakakibara}}, \bibinfo {author}
  {\bibfnamefont {Y.}~\bibnamefont {Takano}}, \bibinfo {author} {\bibfnamefont
  {J.-H.}\ \bibnamefont {Park}}, \bibinfo {author} {\bibfnamefont
  {L.}~\bibnamefont {Balicas}}, \bibinfo {author} {\bibfnamefont
  {N.}~\bibnamefont {Harrison}}, \bibinfo {author} {\bibfnamefont
  {N.}~\bibnamefont {Shitsevalova}}, \bibinfo {author} {\bibfnamefont
  {G.}~\bibnamefont {Balakrishnan}}, \bibinfo {author} {\bibfnamefont {G.~G.}\
  \bibnamefont {Lonzarich}}, \bibinfo {author} {\bibfnamefont {R.~W.}\
  \bibnamefont {Hill}}, \bibinfo {author} {\bibfnamefont {M.}~\bibnamefont
  {Sutherland}},\ and\ \bibinfo {author} {\bibfnamefont {S.~E.}\ \bibnamefont
  {Sebastian}},\ }\bibfield  {title} {\bibinfo {title} {{Fermi surface in the
  absence of a Fermi liquid in the Kondo insulator $\mathrm{SmB}_6$}},\ }\href
  {https://doi.org/10.1038/nphys4295} {\bibfield  {journal} {\bibinfo
  {journal} {Nature Physics}\ }\textbf {\bibinfo {volume} {14}},\ \bibinfo
  {pages} {166} (\bibinfo {year} {2018})}\BibitemShut {NoStop}%
\bibitem [{\citenamefont {Hartstein}\ \emph {et~al.}(2020)\citenamefont
  {Hartstein}, \citenamefont {Liu}, \citenamefont {Hsu}, \citenamefont {Tan},
  \citenamefont {{Ciomaga Hatnean}}, \citenamefont {Balakrishnan},\ and\
  \citenamefont {Sebastian}}]{Hartstein2020}%
  \BibitemOpen
  \bibfield  {author} {\bibinfo {author} {\bibfnamefont {M.}~\bibnamefont
  {Hartstein}}, \bibinfo {author} {\bibfnamefont {H.}~\bibnamefont {Liu}},
  \bibinfo {author} {\bibfnamefont {Y.-T.}\ \bibnamefont {Hsu}}, \bibinfo
  {author} {\bibfnamefont {B.~S.}\ \bibnamefont {Tan}}, \bibinfo {author}
  {\bibfnamefont {M.}~\bibnamefont {{Ciomaga Hatnean}}}, \bibinfo {author}
  {\bibfnamefont {G.}~\bibnamefont {Balakrishnan}},\ and\ \bibinfo {author}
  {\bibfnamefont {S.~E.}\ \bibnamefont {Sebastian}},\ }\bibfield  {title}
  {\bibinfo {title} {{Intrinsic Bulk Quantum Oscillations in a Bulk
  Unconventional Insulator SmB6}},\ }\href
  {https://doi.org/https://doi.org/10.1016/j.isci.2020.101632} {\bibfield
  {journal} {\bibinfo  {journal} {iScience}\ }\textbf {\bibinfo {volume}
  {23}},\ \bibinfo {pages} {101632} (\bibinfo {year} {2020})}\BibitemShut
  {NoStop}%
\bibitem [{\citenamefont {Liu}\ \emph {et~al.}(2018)\citenamefont {Liu},
  \citenamefont {Hartstein}, \citenamefont {Wallace}, \citenamefont {Davies},
  \citenamefont {Hatnean}, \citenamefont {Johannes}, \citenamefont
  {Shitsevalova}, \citenamefont {Balakrishnan},\ and\ \citenamefont
  {Sebastian}}]{Liu2018}%
  \BibitemOpen
  \bibfield  {author} {\bibinfo {author} {\bibfnamefont {H.}~\bibnamefont
  {Liu}}, \bibinfo {author} {\bibfnamefont {M.}~\bibnamefont {Hartstein}},
  \bibinfo {author} {\bibfnamefont {G.~J.}\ \bibnamefont {Wallace}}, \bibinfo
  {author} {\bibfnamefont {A.~J.}\ \bibnamefont {Davies}}, \bibinfo {author}
  {\bibfnamefont {M.~C.}\ \bibnamefont {Hatnean}}, \bibinfo {author}
  {\bibfnamefont {M.~D.}\ \bibnamefont {Johannes}}, \bibinfo {author}
  {\bibfnamefont {N.}~\bibnamefont {Shitsevalova}}, \bibinfo {author}
  {\bibfnamefont {G.}~\bibnamefont {Balakrishnan}},\ and\ \bibinfo {author}
  {\bibfnamefont {S.~E.}\ \bibnamefont {Sebastian}},\ }\bibfield  {title}
  {\bibinfo {title} {{Fermi surfaces in Kondo insulators}},\ }\href
  {https://doi.org/10.1088/1361-648x/aaa522} {\bibfield  {journal} {\bibinfo
  {journal} {Journal of Physics: Condensed Matter}\ }\textbf {\bibinfo {volume}
  {30}},\ \bibinfo {pages} {16LT01} (\bibinfo {year} {2018})}\BibitemShut
  {NoStop}%
\bibitem [{\citenamefont {Xiang}\ \emph {et~al.}(2018)\citenamefont {Xiang},
  \citenamefont {Kasahara}, \citenamefont {Asaba}, \citenamefont {Lawson},
  \citenamefont {Tinsman}, \citenamefont {Chen}, \citenamefont {Sugimoto},
  \citenamefont {Kawaguchi}, \citenamefont {Sato}, \citenamefont {Li},
  \citenamefont {Yao}, \citenamefont {Chen}, \citenamefont {Iga}, \citenamefont
  {Singleton}, \citenamefont {Matsuda},\ and\ \citenamefont {Li}}]{Xiang2018}%
  \BibitemOpen
  \bibfield  {author} {\bibinfo {author} {\bibfnamefont {Z.}~\bibnamefont
  {Xiang}}, \bibinfo {author} {\bibfnamefont {Y.}~\bibnamefont {Kasahara}},
  \bibinfo {author} {\bibfnamefont {T.}~\bibnamefont {Asaba}}, \bibinfo
  {author} {\bibfnamefont {B.}~\bibnamefont {Lawson}}, \bibinfo {author}
  {\bibfnamefont {C.}~\bibnamefont {Tinsman}}, \bibinfo {author} {\bibfnamefont
  {L.}~\bibnamefont {Chen}}, \bibinfo {author} {\bibfnamefont {K.}~\bibnamefont
  {Sugimoto}}, \bibinfo {author} {\bibfnamefont {S.}~\bibnamefont {Kawaguchi}},
  \bibinfo {author} {\bibfnamefont {Y.}~\bibnamefont {Sato}}, \bibinfo {author}
  {\bibfnamefont {G.}~\bibnamefont {Li}}, \bibinfo {author} {\bibfnamefont
  {S.}~\bibnamefont {Yao}}, \bibinfo {author} {\bibfnamefont {Y.~L.}\
  \bibnamefont {Chen}}, \bibinfo {author} {\bibfnamefont {F.}~\bibnamefont
  {Iga}}, \bibinfo {author} {\bibfnamefont {J.}~\bibnamefont {Singleton}},
  \bibinfo {author} {\bibfnamefont {Y.}~\bibnamefont {Matsuda}},\ and\ \bibinfo
  {author} {\bibfnamefont {L.}~\bibnamefont {Li}},\ }\bibfield  {title}
  {\bibinfo {title} {{Quantum oscillations of electrical resistivity in an
  insulator}},\ }\href {https://doi.org/10.1126/science.aap9607} {\bibfield
  {journal} {\bibinfo  {journal} {Science}\ }\textbf {\bibinfo {volume}
  {362}},\ \bibinfo {pages} {65} (\bibinfo {year} {2018})}\BibitemShut
  {NoStop}%
\bibitem [{\citenamefont {Liu}\ \emph {et~al.}(2022)\citenamefont {Liu},
  \citenamefont {Hickey}, \citenamefont {Hartstein}, \citenamefont {Davies},
  \citenamefont {Eaton}, \citenamefont {Elvin}, \citenamefont {Polyakov},
  \citenamefont {Vu}, \citenamefont {Wichitwechkarn}, \citenamefont
  {F{\"o}rster}, \citenamefont {Wosnitza}, \citenamefont {Murphy},
  \citenamefont {Shitsevalova}, \citenamefont {Johannes}, \citenamefont
  {Hatnean}, \citenamefont {Balakrishnan}, \citenamefont {Lonzarich},\ and\
  \citenamefont {Sebastian}}]{Liu2022}%
  \BibitemOpen
  \bibfield  {author} {\bibinfo {author} {\bibfnamefont {H.}~\bibnamefont
  {Liu}}, \bibinfo {author} {\bibfnamefont {A.~J.}\ \bibnamefont {Hickey}},
  \bibinfo {author} {\bibfnamefont {M.}~\bibnamefont {Hartstein}}, \bibinfo
  {author} {\bibfnamefont {A.~J.}\ \bibnamefont {Davies}}, \bibinfo {author}
  {\bibfnamefont {A.~G.}\ \bibnamefont {Eaton}}, \bibinfo {author}
  {\bibfnamefont {T.}~\bibnamefont {Elvin}}, \bibinfo {author} {\bibfnamefont
  {E.}~\bibnamefont {Polyakov}}, \bibinfo {author} {\bibfnamefont {T.~H.}\
  \bibnamefont {Vu}}, \bibinfo {author} {\bibfnamefont {V.}~\bibnamefont
  {Wichitwechkarn}}, \bibinfo {author} {\bibfnamefont {T.}~\bibnamefont
  {F{\"o}rster}}, \bibinfo {author} {\bibfnamefont {J.}~\bibnamefont
  {Wosnitza}}, \bibinfo {author} {\bibfnamefont {T.~P.}\ \bibnamefont
  {Murphy}}, \bibinfo {author} {\bibfnamefont {N.}~\bibnamefont
  {Shitsevalova}}, \bibinfo {author} {\bibfnamefont {M.~D.}\ \bibnamefont
  {Johannes}}, \bibinfo {author} {\bibfnamefont {M.~C.}\ \bibnamefont
  {Hatnean}}, \bibinfo {author} {\bibfnamefont {G.}~\bibnamefont
  {Balakrishnan}}, \bibinfo {author} {\bibfnamefont {G.~G.}\ \bibnamefont
  {Lonzarich}},\ and\ \bibinfo {author} {\bibfnamefont {S.~E.}\ \bibnamefont
  {Sebastian}},\ }\bibfield  {title} {\bibinfo {title} {{f-electron hybridised
  Fermi surface in magnetic field-induced metallic YbB12}},\ }\href
  {https://doi.org/10.1038/s41535-021-00413-7} {\bibfield  {journal} {\bibinfo
  {journal} {npj Quantum Materials}\ }\textbf {\bibinfo {volume} {7}},\
  \bibinfo {pages} {12} (\bibinfo {year} {2022})}\BibitemShut {NoStop}%
\bibitem [{\citenamefont {Lifshitz}\ and\ \citenamefont
  {Kosevich}(1956)}]{Lifshitz1956}%
  \BibitemOpen
  \bibfield  {author} {\bibinfo {author} {\bibfnamefont {I.}~\bibnamefont
  {Lifshitz}}\ and\ \bibinfo {author} {\bibfnamefont {A.}~\bibnamefont
  {Kosevich}},\ }\bibfield  {title} {\bibinfo {title} {{Theory of Magnetic
  Susceptibility in Metals at Low Temperature}},\ }\href@noop {} {\bibfield
  {journal} {\bibinfo  {journal} {Soviet Phys. JETP}\ }\textbf {\bibinfo
  {volume} {2}},\ \bibinfo {pages} {636} (\bibinfo {year} {1956})}\BibitemShut
  {NoStop}%
\bibitem [{\citenamefont {Shoenberg}(1984)}]{Shoenberg1984}%
  \BibitemOpen
  \bibfield  {author} {\bibinfo {author} {\bibfnamefont {D.}~\bibnamefont
  {Shoenberg}},\ }\href {https://doi.org/10.1017/CBO9780511897870} {\emph
  {\bibinfo {title} {{Magnetic Oscillations in Metals}}}},\ Cambridge
  Monographs on Physics\ (\bibinfo  {publisher} {Cambridge University Press},\
  \bibinfo {year} {1984})\BibitemShut {NoStop}%
\bibitem [{\citenamefont {Baskaran}(2015)}]{Baskaran2015}%
  \BibitemOpen
  \bibfield  {author} {\bibinfo {author} {\bibfnamefont {G.}~\bibnamefont
  {Baskaran}},\ }\href@noop {} {\bibinfo {title} {{Majorana Fermi Sea in
  Insulating SmB6: A proposal and a Theory of Quantum Oscillations in Kondo
  Insulators}}} (\bibinfo {year} {2015}),\ \Eprint
  {https://arxiv.org/abs/1507.03477} {arXiv:1507.03477 [cond-mat.str-el]}
  \BibitemShut {NoStop}%
\bibitem [{\citenamefont {Erten}\ \emph {et~al.}(2016)\citenamefont {Erten},
  \citenamefont {Ghaemi},\ and\ \citenamefont {Coleman}}]{Erten2016}%
  \BibitemOpen
  \bibfield  {author} {\bibinfo {author} {\bibfnamefont {O.}~\bibnamefont
  {Erten}}, \bibinfo {author} {\bibfnamefont {P.}~\bibnamefont {Ghaemi}},\ and\
  \bibinfo {author} {\bibfnamefont {P.}~\bibnamefont {Coleman}},\ }\bibfield
  {title} {\bibinfo {title} {{Kondo Breakdown and Quantum Oscillations in
  ${\mathrm{SmB}}_{6}$}},\ }\href
  {https://doi.org/10.1103/PhysRevLett.116.046403} {\bibfield  {journal}
  {\bibinfo  {journal} {Phys. Rev. Lett.}\ }\textbf {\bibinfo {volume} {116}},\
  \bibinfo {pages} {046403} (\bibinfo {year} {2016})}\BibitemShut {NoStop}%
\bibitem [{\citenamefont {Knolle}\ and\ \citenamefont
  {Cooper}(2017{\natexlab{a}})}]{Knolle2017a}%
  \BibitemOpen
  \bibfield  {author} {\bibinfo {author} {\bibfnamefont {J.}~\bibnamefont
  {Knolle}}\ and\ \bibinfo {author} {\bibfnamefont {N.~R.}\ \bibnamefont
  {Cooper}},\ }\bibfield  {title} {\bibinfo {title} {{Excitons in topological
  Kondo insulators: Theory of thermodynamic and transport anomalies in
  ${\mathrm{SmB}}_{6}$}},\ }\href
  {https://doi.org/10.1103/PhysRevLett.118.096604} {\bibfield  {journal}
  {\bibinfo  {journal} {Phys. Rev. Lett.}\ }\textbf {\bibinfo {volume} {118}},\
  \bibinfo {pages} {096604} (\bibinfo {year} {2017}{\natexlab{a}})}\BibitemShut
  {NoStop}%
\bibitem [{\citenamefont {Erten}\ \emph {et~al.}(2017)\citenamefont {Erten},
  \citenamefont {Chang}, \citenamefont {Coleman},\ and\ \citenamefont
  {Tsvelik}}]{Erten2017}%
  \BibitemOpen
  \bibfield  {author} {\bibinfo {author} {\bibfnamefont {O.}~\bibnamefont
  {Erten}}, \bibinfo {author} {\bibfnamefont {P.-Y.}\ \bibnamefont {Chang}},
  \bibinfo {author} {\bibfnamefont {P.}~\bibnamefont {Coleman}},\ and\ \bibinfo
  {author} {\bibfnamefont {A.~M.}\ \bibnamefont {Tsvelik}},\ }\bibfield
  {title} {\bibinfo {title} {Skyrme insulators: Insulators at the brink of
  superconductivity},\ }\href {https://doi.org/10.1103/PhysRevLett.119.057603}
  {\bibfield  {journal} {\bibinfo  {journal} {Phys. Rev. Lett.}\ }\textbf
  {\bibinfo {volume} {119}},\ \bibinfo {pages} {057603} (\bibinfo {year}
  {2017})}\BibitemShut {NoStop}%
\bibitem [{\citenamefont {Riseborough}\ and\ \citenamefont
  {Fisk}(2017)}]{Riseborough2017}%
  \BibitemOpen
  \bibfield  {author} {\bibinfo {author} {\bibfnamefont {P.~S.}\ \bibnamefont
  {Riseborough}}\ and\ \bibinfo {author} {\bibfnamefont {Z.}~\bibnamefont
  {Fisk}},\ }\bibfield  {title} {\bibinfo {title} {{Critical examination of
  quantum oscillations in ${\mathrm{SmB}}_{6}$}},\ }\href
  {https://doi.org/10.1103/PhysRevB.96.195122} {\bibfield  {journal} {\bibinfo
  {journal} {Phys. Rev. B}\ }\textbf {\bibinfo {volume} {96}},\ \bibinfo
  {pages} {195122} (\bibinfo {year} {2017})}\BibitemShut {NoStop}%
\bibitem [{\citenamefont {Sodemann}\ \emph {et~al.}(2018)\citenamefont
  {Sodemann}, \citenamefont {Chowdhury},\ and\ \citenamefont
  {Senthil}}]{Sodemann2018}%
  \BibitemOpen
  \bibfield  {author} {\bibinfo {author} {\bibfnamefont {I.}~\bibnamefont
  {Sodemann}}, \bibinfo {author} {\bibfnamefont {D.}~\bibnamefont
  {Chowdhury}},\ and\ \bibinfo {author} {\bibfnamefont {T.}~\bibnamefont
  {Senthil}},\ }\bibfield  {title} {\bibinfo {title} {{Quantum oscillations in
  insulators with neutral Fermi surfaces}},\ }\href
  {https://doi.org/10.1103/PhysRevB.97.045152} {\bibfield  {journal} {\bibinfo
  {journal} {Phys. Rev. B}\ }\textbf {\bibinfo {volume} {97}},\ \bibinfo
  {pages} {045152} (\bibinfo {year} {2018})}\BibitemShut {NoStop}%
\bibitem [{\citenamefont {Chowdhury}\ \emph {et~al.}(2018)\citenamefont
  {Chowdhury}, \citenamefont {Sodemann},\ and\ \citenamefont
  {Senthil}}]{Chowdhury2018}%
  \BibitemOpen
  \bibfield  {author} {\bibinfo {author} {\bibfnamefont {D.}~\bibnamefont
  {Chowdhury}}, \bibinfo {author} {\bibfnamefont {I.}~\bibnamefont
  {Sodemann}},\ and\ \bibinfo {author} {\bibfnamefont {T.}~\bibnamefont
  {Senthil}},\ }\bibfield  {title} {\bibinfo {title} {Mixed-valence insulators
  with neutral fermi surfaces},\ }\href
  {https://doi.org/10.1038/s41467-018-04163-2} {\bibfield  {journal} {\bibinfo
  {journal} {Nature Communications}\ }\textbf {\bibinfo {volume} {9}},\
  \bibinfo {pages} {1766} (\bibinfo {year} {2018})}\BibitemShut {NoStop}%
\bibitem [{\citenamefont {Peters}\ \emph {et~al.}(2019)\citenamefont {Peters},
  \citenamefont {Yoshida},\ and\ \citenamefont {Kawakami}}]{Peters2019}%
  \BibitemOpen
  \bibfield  {author} {\bibinfo {author} {\bibfnamefont {R.}~\bibnamefont
  {Peters}}, \bibinfo {author} {\bibfnamefont {T.}~\bibnamefont {Yoshida}},\
  and\ \bibinfo {author} {\bibfnamefont {N.}~\bibnamefont {Kawakami}},\
  }\bibfield  {title} {\bibinfo {title} {{Quantum oscillations in strongly
  correlated topological Kondo insulators}},\ }\href
  {https://doi.org/10.1103/PhysRevB.100.085124} {\bibfield  {journal} {\bibinfo
   {journal} {Phys. Rev. B}\ }\textbf {\bibinfo {volume} {100}},\ \bibinfo
  {pages} {085124} (\bibinfo {year} {2019})}\BibitemShut {NoStop}%
\bibitem [{\citenamefont {Lu}\ \emph {et~al.}(2020)\citenamefont {Lu},
  \citenamefont {Chou}, \citenamefont {Chung}, \citenamefont {Lee},\ and\
  \citenamefont {Mou}}]{Lu2020}%
  \BibitemOpen
  \bibfield  {author} {\bibinfo {author} {\bibfnamefont {Y.-W.}\ \bibnamefont
  {Lu}}, \bibinfo {author} {\bibfnamefont {P.-H.}\ \bibnamefont {Chou}},
  \bibinfo {author} {\bibfnamefont {C.-H.}\ \bibnamefont {Chung}}, \bibinfo
  {author} {\bibfnamefont {T.-K.}\ \bibnamefont {Lee}},\ and\ \bibinfo {author}
  {\bibfnamefont {C.-Y.}\ \bibnamefont {Mou}},\ }\bibfield  {title} {\bibinfo
  {title} {{Enhanced quantum oscillations in Kondo insulators}},\ }\href
  {https://doi.org/10.1103/PhysRevB.101.115102} {\bibfield  {journal} {\bibinfo
   {journal} {Phys. Rev. B}\ }\textbf {\bibinfo {volume} {101}},\ \bibinfo
  {pages} {115102} (\bibinfo {year} {2020})}\BibitemShut {NoStop}%
\bibitem [{\citenamefont {Varma}(2020)}]{Varma2020}%
  \BibitemOpen
  \bibfield  {author} {\bibinfo {author} {\bibfnamefont {C.~M.}\ \bibnamefont
  {Varma}},\ }\bibfield  {title} {\bibinfo {title} {Majoranas in mixed-valence
  insulators},\ }\href {https://doi.org/10.1103/PhysRevB.102.155145} {\bibfield
   {journal} {\bibinfo  {journal} {Phys. Rev. B}\ }\textbf {\bibinfo {volume}
  {102}},\ \bibinfo {pages} {155145} (\bibinfo {year} {2020})}\BibitemShut
  {NoStop}%
\bibitem [{\citenamefont {Knolle}\ and\ \citenamefont
  {Cooper}(2015)}]{Knolle2015}%
  \BibitemOpen
  \bibfield  {author} {\bibinfo {author} {\bibfnamefont {J.}~\bibnamefont
  {Knolle}}\ and\ \bibinfo {author} {\bibfnamefont {N.~R.}\ \bibnamefont
  {Cooper}},\ }\bibfield  {title} {\bibinfo {title} {{Quantum Oscillations
  without a Fermi Surface and the Anomalous de Haas--van Alphen Effect}},\
  }\href {https://doi.org/10.1103/PhysRevLett.115.146401} {\bibfield  {journal}
  {\bibinfo  {journal} {Phys. Rev. Lett.}\ }\textbf {\bibinfo {volume} {115}},\
  \bibinfo {pages} {146401} (\bibinfo {year} {2015})}\BibitemShut {NoStop}%
\bibitem [{\citenamefont {Zhang}\ \emph {et~al.}(2016)\citenamefont {Zhang},
  \citenamefont {Song},\ and\ \citenamefont {Wang}}]{Zhang2016}%
  \BibitemOpen
  \bibfield  {author} {\bibinfo {author} {\bibfnamefont {L.}~\bibnamefont
  {Zhang}}, \bibinfo {author} {\bibfnamefont {X.-Y.}\ \bibnamefont {Song}},\
  and\ \bibinfo {author} {\bibfnamefont {F.}~\bibnamefont {Wang}},\ }\bibfield
  {title} {\bibinfo {title} {{Quantum Oscillation in Narrow-Gap Topological
  Insulators}},\ }\href {https://doi.org/10.1103/PhysRevLett.116.046404}
  {\bibfield  {journal} {\bibinfo  {journal} {Phys. Rev. Lett.}\ }\textbf
  {\bibinfo {volume} {116}},\ \bibinfo {pages} {046404} (\bibinfo {year}
  {2016})}\BibitemShut {NoStop}%
\bibitem [{\citenamefont {Pal}\ \emph {et~al.}(2016)\citenamefont {Pal},
  \citenamefont {Pi\'echon}, \citenamefont {Fuchs}, \citenamefont {Goerbig},\
  and\ \citenamefont {Montambaux}}]{Pal2016}%
  \BibitemOpen
  \bibfield  {author} {\bibinfo {author} {\bibfnamefont {H.~K.}\ \bibnamefont
  {Pal}}, \bibinfo {author} {\bibfnamefont {F.}~\bibnamefont {Pi\'echon}},
  \bibinfo {author} {\bibfnamefont {J.-N.}\ \bibnamefont {Fuchs}}, \bibinfo
  {author} {\bibfnamefont {M.}~\bibnamefont {Goerbig}},\ and\ \bibinfo {author}
  {\bibfnamefont {G.}~\bibnamefont {Montambaux}},\ }\bibfield  {title}
  {\bibinfo {title} {Chemical potential asymmetry and quantum oscillations in
  insulators},\ }\href {https://doi.org/10.1103/PhysRevB.94.125140} {\bibfield
  {journal} {\bibinfo  {journal} {Phys. Rev. B}\ }\textbf {\bibinfo {volume}
  {94}},\ \bibinfo {pages} {125140} (\bibinfo {year} {2016})}\BibitemShut
  {NoStop}%
\bibitem [{\citenamefont {Pal}(2017{\natexlab{a}})}]{Pal2017a}%
  \BibitemOpen
  \bibfield  {author} {\bibinfo {author} {\bibfnamefont {H.~K.}\ \bibnamefont
  {Pal}},\ }\bibfield  {title} {\bibinfo {title} {Quantum oscillations from
  inside the fermi sea},\ }\href {https://doi.org/10.1103/PhysRevB.95.085111}
  {\bibfield  {journal} {\bibinfo  {journal} {Phys. Rev. B}\ }\textbf {\bibinfo
  {volume} {95}},\ \bibinfo {pages} {085111} (\bibinfo {year}
  {2017}{\natexlab{a}})}\BibitemShut {NoStop}%
\bibitem [{\citenamefont {Pal}(2017{\natexlab{b}})}]{Pal2017b}%
  \BibitemOpen
  \bibfield  {author} {\bibinfo {author} {\bibfnamefont {H.~K.}\ \bibnamefont
  {Pal}},\ }\bibfield  {title} {\bibinfo {title} {Unusual frequency of quantum
  oscillations in strongly particle-hole asymmetric insulators},\ }\href
  {https://doi.org/10.1103/PhysRevB.96.235121} {\bibfield  {journal} {\bibinfo
  {journal} {Phys. Rev. B}\ }\textbf {\bibinfo {volume} {96}},\ \bibinfo
  {pages} {235121} (\bibinfo {year} {2017}{\natexlab{b}})}\BibitemShut
  {NoStop}%
\bibitem [{\citenamefont {Knolle}\ and\ \citenamefont
  {Cooper}(2017{\natexlab{b}})}]{Knolle2017b}%
  \BibitemOpen
  \bibfield  {author} {\bibinfo {author} {\bibfnamefont {J.}~\bibnamefont
  {Knolle}}\ and\ \bibinfo {author} {\bibfnamefont {N.~R.}\ \bibnamefont
  {Cooper}},\ }\bibfield  {title} {\bibinfo {title} {{Anomalous de Haas--van
  Alphen Effect in $\mathrm{InAs}/\mathrm{GaSb}$ Quantum Wells}},\ }\href
  {https://doi.org/10.1103/PhysRevLett.118.176801} {\bibfield  {journal}
  {\bibinfo  {journal} {Phys. Rev. Lett.}\ }\textbf {\bibinfo {volume} {118}},\
  \bibinfo {pages} {176801} (\bibinfo {year} {2017}{\natexlab{b}})}\BibitemShut
  {NoStop}%
\bibitem [{\citenamefont {Shen}\ and\ \citenamefont {Fu}(2018)}]{Shen2018}%
  \BibitemOpen
  \bibfield  {author} {\bibinfo {author} {\bibfnamefont {H.}~\bibnamefont
  {Shen}}\ and\ \bibinfo {author} {\bibfnamefont {L.}~\bibnamefont {Fu}},\
  }\bibfield  {title} {\bibinfo {title} {Quantum oscillation from in-gap states
  and a non-{H}ermitian {L}andau level problem},\ }\href
  {https://doi.org/10.1103/PhysRevLett.121.026403} {\bibfield  {journal}
  {\bibinfo  {journal} {Phys. Rev. Lett.}\ }\textbf {\bibinfo {volume} {121}},\
  \bibinfo {pages} {026403} (\bibinfo {year} {2018})}\BibitemShut {NoStop}%
\bibitem [{\citenamefont {Skinner}(2019)}]{Skinner2019}%
  \BibitemOpen
  \bibfield  {author} {\bibinfo {author} {\bibfnamefont {B.}~\bibnamefont
  {Skinner}},\ }\bibfield  {title} {\bibinfo {title} {Properties of the donor
  impurity band in mixed valence insulators},\ }\href
  {https://doi.org/10.1103/PhysRevMaterials.3.104601} {\bibfield  {journal}
  {\bibinfo  {journal} {Phys. Rev. Materials}\ }\textbf {\bibinfo {volume}
  {3}},\ \bibinfo {pages} {104601} (\bibinfo {year} {2019})}\BibitemShut
  {NoStop}%
\bibitem [{\citenamefont {Lee}(2021)}]{Lee2021}%
  \BibitemOpen
  \bibfield  {author} {\bibinfo {author} {\bibfnamefont {P.~A.}\ \bibnamefont
  {Lee}},\ }\bibfield  {title} {\bibinfo {title} {{Quantum oscillations in the
  activated conductivity in excitonic insulators: Possible application to
  monolayer ${\mathrm{WTe}}_{2}$}},\ }\href
  {https://doi.org/10.1103/PhysRevB.103.L041101} {\bibfield  {journal}
  {\bibinfo  {journal} {Phys. Rev. B}\ }\textbf {\bibinfo {volume} {103}},\
  \bibinfo {pages} {L041101} (\bibinfo {year} {2021})}\BibitemShut {NoStop}%
\bibitem [{\citenamefont {He}\ and\ \citenamefont {Lee}(2021)}]{He2021}%
  \BibitemOpen
  \bibfield  {author} {\bibinfo {author} {\bibfnamefont {W.-Y.}\ \bibnamefont
  {He}}\ and\ \bibinfo {author} {\bibfnamefont {P.~A.}\ \bibnamefont {Lee}},\
  }\bibfield  {title} {\bibinfo {title} {{Quantum oscillation of thermally
  activated conductivity in a monolayer ${\mathrm{WTe}}_{2}$-like excitonic
  insulator}},\ }\href {https://doi.org/10.1103/PhysRevB.104.L041110}
  {\bibfield  {journal} {\bibinfo  {journal} {Phys. Rev. B}\ }\textbf {\bibinfo
  {volume} {104}},\ \bibinfo {pages} {L041110} (\bibinfo {year}
  {2021})}\BibitemShut {NoStop}%
\bibitem [{\citenamefont {Allocca}\ and\ \citenamefont
  {Cooper}(2022)}]{Allocca2022}%
  \BibitemOpen
  \bibfield  {author} {\bibinfo {author} {\bibfnamefont {A.~A.}\ \bibnamefont
  {Allocca}}\ and\ \bibinfo {author} {\bibfnamefont {N.~R.}\ \bibnamefont
  {Cooper}},\ }\bibfield  {title} {\bibinfo {title} {{Quantum oscillations in
  interaction-driven insulators}},\ }\href
  {https://doi.org/10.21468/SciPostPhys.12.4.123} {\bibfield  {journal}
  {\bibinfo  {journal} {SciPost Phys.}\ }\textbf {\bibinfo {volume} {12}},\
  \bibinfo {pages} {123} (\bibinfo {year} {2022})}\BibitemShut {NoStop}%
\bibitem [{\citenamefont {Panda}\ \emph {et~al.}(2022)\citenamefont {Panda},
  \citenamefont {Banerjee},\ and\ \citenamefont {Randeria}}]{Randeria2023}%
  \BibitemOpen
  \bibfield  {author} {\bibinfo {author} {\bibfnamefont {A.}~\bibnamefont
  {Panda}}, \bibinfo {author} {\bibfnamefont {S.}~\bibnamefont {Banerjee}},\
  and\ \bibinfo {author} {\bibfnamefont {M.}~\bibnamefont {Randeria}},\
  }\bibfield  {title} {\bibinfo {title} {Quantum oscillations in the
  magnetization and density of states of insulators},\ }\href
  {https://doi.org/10.1073/pnas.2208373119} {\bibfield  {journal} {\bibinfo
  {journal} {Proceedings of the National Academy of Sciences}\ }\textbf
  {\bibinfo {volume} {119}},\ \bibinfo {pages} {e2208373119} (\bibinfo {year}
  {2022})}\BibitemShut {NoStop}%
\bibitem [{\citenamefont {Julian}(2023)}]{Julian2023}%
  \BibitemOpen
  \bibfield  {author} {\bibinfo {author} {\bibfnamefont {S.~R.}\ \bibnamefont
  {Julian}},\ }\href {https://doi.org/10.48550/ARXIV.2301.05366} {\bibinfo
  {title} {de haas van alphen oscillations in hybridization-gap insulators as a
  sudden change in the diamagnetic moment of landau levels}} (\bibinfo {year}
  {2023})\BibitemShut {NoStop}%
\bibitem [{\citenamefont {Vaks}\ \emph {et~al.}(1962)\citenamefont {Vaks},
  \citenamefont {Galitskii},\ and\ \citenamefont {Larkin}}]{Vaks1962}%
  \BibitemOpen
  \bibfield  {author} {\bibinfo {author} {\bibfnamefont {V.~G.}\ \bibnamefont
  {Vaks}}, \bibinfo {author} {\bibfnamefont {V.~M.}\ \bibnamefont
  {Galitskii}},\ and\ \bibinfo {author} {\bibfnamefont {A.~I.}\ \bibnamefont
  {Larkin}},\ }\bibfield  {title} {\bibinfo {title} {Collective excitations in
  a superconductor},\ }\href@noop {} {\bibfield  {journal} {\bibinfo  {journal}
  {Soviet Phys. JETP}\ }\textbf {\bibinfo {volume} {41}},\ \bibinfo {pages}
  {1655} (\bibinfo {year} {1962})}\BibitemShut {NoStop}%
\bibitem [{\citenamefont {Kos}\ \emph {et~al.}(2004)\citenamefont {Kos},
  \citenamefont {Millis},\ and\ \citenamefont {Larkin}}]{Kos2004}%
  \BibitemOpen
  \bibfield  {author} {\bibinfo {author} {\bibfnamefont {S.}~\bibnamefont
  {Kos}}, \bibinfo {author} {\bibfnamefont {A.~J.}\ \bibnamefont {Millis}},\
  and\ \bibinfo {author} {\bibfnamefont {A.~I.}\ \bibnamefont {Larkin}},\
  }\bibfield  {title} {\bibinfo {title} {{Gaussian fluctuation corrections to
  the BCS mean-field gap amplitude at zero temperature}},\ }\href
  {https://doi.org/10.1103/PhysRevB.70.214531} {\bibfield  {journal} {\bibinfo
  {journal} {Phys. Rev. B}\ }\textbf {\bibinfo {volume} {70}},\ \bibinfo
  {pages} {214531} (\bibinfo {year} {2004})}\BibitemShut {NoStop}%
\bibitem [{\citenamefont {Hoyer}\ and\ \citenamefont
  {Schmalian}(2018)}]{Hoyer2018}%
  \BibitemOpen
  \bibfield  {author} {\bibinfo {author} {\bibfnamefont {M.}~\bibnamefont
  {Hoyer}}\ and\ \bibinfo {author} {\bibfnamefont {J.}~\bibnamefont
  {Schmalian}},\ }\bibfield  {title} {\bibinfo {title} {Role of fluctuations
  for density-wave instabilities: Failure of the mean-field description},\
  }\href {https://doi.org/10.1103/PhysRevB.97.224423} {\bibfield  {journal}
  {\bibinfo  {journal} {Phys. Rev. B}\ }\textbf {\bibinfo {volume} {97}},\
  \bibinfo {pages} {224423} (\bibinfo {year} {2018})}\BibitemShut {NoStop}%
\bibitem [{Sup()}]{Supplement}%
  \BibitemOpen
  \href@noop {} {}\bibinfo {note} {See Supplemental Material for the detailed
  effect of the magnetic field and calculations of free energies and related
  intermediate quantities.}\BibitemShut {Stop}%
\bibitem [{Note1()}]{Note1}%
  \BibitemOpen
  \bibinfo {note} {Much of this numerical analysis is done using the Julia
  programming language~\cite {Bezanson2017}.}\BibitemShut {Stop}%
\bibitem [{Note2()}]{Note2}%
  \BibitemOpen
  \bibinfo {note} {We find that the results for different $T\leq \Delta _0/2$
  are nearly indistinguishable, so these numerical results are a very good
  reflection of the $T=0$ behavior of the system.}\BibitemShut {Stop}%
\bibitem [{\citenamefont {Allocca}\ and\ \citenamefont
  {Cooper}(2021)}]{Allocca2021}%
  \BibitemOpen
  \bibfield  {author} {\bibinfo {author} {\bibfnamefont {A.~A.}\ \bibnamefont
  {Allocca}}\ and\ \bibinfo {author} {\bibfnamefont {N.~R.}\ \bibnamefont
  {Cooper}},\ }\bibfield  {title} {\bibinfo {title} {Low-frequency quantum
  oscillations from interactions in layered metals},\ }\href
  {https://doi.org/10.1103/PhysRevResearch.3.L042009} {\bibfield  {journal}
  {\bibinfo  {journal} {Phys. Rev. Research}\ }\textbf {\bibinfo {volume}
  {3}},\ \bibinfo {pages} {L042009} (\bibinfo {year} {2021})}\BibitemShut
  {NoStop}%
\bibitem [{\citenamefont {Leeb}\ and\ \citenamefont {Knolle}()}]{Leeb2023}%
  \BibitemOpen
  \bibfield  {author} {\bibinfo {author} {\bibfnamefont {V.}~\bibnamefont
  {Leeb}}\ and\ \bibinfo {author} {\bibfnamefont {J.}~\bibnamefont {Knolle}},\
  }\bibfield  {title} {\bibinfo {title} {Quantum oscillations in a doped {M}ott
  insulator beyond {O}nsager's relation},\ }\href
  {https://doi.org/10.48550/arXiv.2301.08685} {\bibinfo  {journal}
  {arXiv:2301.08685}\ }\BibitemShut {NoStop}%
\bibitem [{\citenamefont {Le}\ \emph {et~al.}(2015)\citenamefont {Le},
  \citenamefont {Barinov}, \citenamefont {Preciado}, \citenamefont {Isarraraz},
  \citenamefont {Tanabe}, \citenamefont {Komesu}, \citenamefont {Troha},
  \citenamefont {Bartels}, \citenamefont {Rahman},\ and\ \citenamefont
  {Dowben}}]{Le2015}%
  \BibitemOpen
\bibfield  {journal} {  }\bibfield  {author} {\bibinfo {author} {\bibfnamefont
  {D.}~\bibnamefont {Le}}, \bibinfo {author} {\bibfnamefont {A.}~\bibnamefont
  {Barinov}}, \bibinfo {author} {\bibfnamefont {E.}~\bibnamefont {Preciado}},
  \bibinfo {author} {\bibfnamefont {M.}~\bibnamefont {Isarraraz}}, \bibinfo
  {author} {\bibfnamefont {I.}~\bibnamefont {Tanabe}}, \bibinfo {author}
  {\bibfnamefont {T.}~\bibnamefont {Komesu}}, \bibinfo {author} {\bibfnamefont
  {C.}~\bibnamefont {Troha}}, \bibinfo {author} {\bibfnamefont
  {L.}~\bibnamefont {Bartels}}, \bibinfo {author} {\bibfnamefont {T.~S.}\
  \bibnamefont {Rahman}},\ and\ \bibinfo {author} {\bibfnamefont {P.~A.}\
  \bibnamefont {Dowben}},\ }\bibfield  {title} {\bibinfo {title} {Spin–orbit
  coupling in the band structure of monolayer wse2},\ }\href
  {https://doi.org/10.1088/0953-8984/27/18/182201} {\bibfield  {journal}
  {\bibinfo  {journal} {Journal of Physics: Condensed Matter}\ }\textbf
  {\bibinfo {volume} {27}},\ \bibinfo {pages} {182201} (\bibinfo {year}
  {2015})}\BibitemShut {NoStop}%
\bibitem [{\citenamefont {Bezanson}\ \emph {et~al.}(2017)\citenamefont
  {Bezanson}, \citenamefont {Edelman}, \citenamefont {Karpinski},\ and\
  \citenamefont {Shah}}]{Bezanson2017}%
  \BibitemOpen
  \bibfield  {author} {\bibinfo {author} {\bibfnamefont {J.}~\bibnamefont
  {Bezanson}}, \bibinfo {author} {\bibfnamefont {A.}~\bibnamefont {Edelman}},
  \bibinfo {author} {\bibfnamefont {S.}~\bibnamefont {Karpinski}},\ and\
  \bibinfo {author} {\bibfnamefont {V.~B.}\ \bibnamefont {Shah}},\ }\bibfield
  {title} {\bibinfo {title} {Julia: A fresh approach to numerical computing},\
  }\href {https://doi.org/10.1137/141000671} {\bibfield  {journal} {\bibinfo
  {journal} {SIAM review}\ }\textbf {\bibinfo {volume} {59}},\ \bibinfo {pages}
  {65} (\bibinfo {year} {2017})}\BibitemShut {NoStop}%
\end{thebibliography}%
